\pdfoutput=1
\documentclass[a4paper,fleqn]{cas-sc}

\usepackage{flafter}   
\usepackage[numbers]{natbib}
\usepackage{epstopdf}
\usepackage{subcaption}
\usepackage{booktabs}
\usepackage{multirow}
\usepackage{placeins}
\usepackage{enumitem}
\usepackage{amsmath}
\usepackage{tikz}
\newcommand{\deck}[1]{\texttt{#1}}
\usetikzlibrary{shapes.geometric,shapes.misc,arrows.meta,positioning,calc,fit,backgrounds,patterns}

\DeclareMathOperator{\tr}{tr}

\newcommand{\vect}[1]{\boldsymbol{#1}}
\newcommand{\mat}[1]{\mathbf{#1}}
\newcommand{\R}{\mathbb{R}}
\newcommand{\T}{^{\mathsf{T}}}
\newcommand{\Kg}{\mat{K}_{\mathrm{g}}}
\newcommand{\Kref}{k_{\mathrm{ref}}}

\newcommand{\Emin}{\varepsilon_{E}}

\begin{document}
\let\WriteBookmarks\relax
\def\floatpagepagefraction{1}
\def\textpagefraction{.001}
\shorttitle{Formulation-dependent buckling of stiffened shell panels}
\shortauthors{L.~Wang et~al.}

\title [mode = title]{Formulation-dependent linearized buckling of stiffened shell panels: role of the stress stiffness operator}
\tnotemark[1]

\tnotetext[1]{This research is supported by the National Natural Science Foundation of China (Grant No.~51405397, 51675450), 
the Sichuan Provincial Science and Technology Program (Grant No.~2023YFG0182),
the Southwest Jiaotong University Multidisciplinary Research Fund (No.~2682025ZD005), 
the Major Science and Technology Projects in Sichuan, China (grant number: 2023ZDZX0009),
and the Science and Technology Development Program Project of China Railway Eryuan Engineering Group Co., Ltd. (grant number: KSNQ253005).
}

\author[1,2]{Lifeng Wang}
\credit{Methodology, Software, Investigation, Validation, Writing -- original draft, Funding acquisition}

\author[1]{Zhongli Qiu}
\credit{Data curation, Software, Writing -- original draft}

\affiliation[1]{organization={School of Mechanical Engineering, Southwest Jiaotong University},
city={Chengdu},
citysep={},
postcode={610031},
country={China}}

\affiliation[2]{organization={China Railway Eryuan Engineering Group Co., Ltd.},
city={Chengdu},
citysep={},
postcode={610031},
country={China}}

\author[1]{Xuanhao Cheng}
\credit{Software, Validation, Writing -- original draft}

\affiliation[3]{organization={Technology and Equipment of Rail Transit Operation and Maintenance Key Laboratory of Sichuan Province, Southwest Jiaotong University},
city={Chengdu},
citysep={},
postcode={610031},
country={China}}

\author[1,3,4]{Run Du}[orcid=0000-0002-6614-2752]
\cormark[1]
\ead{rdu@swjtu.edu.cn}
\ead[url]{https://faculty.swjtu.edu.cn/durun}
\credit{Conceptualization, Software, Writing -- review \& editing, Funding acquisition}

\author[1,4]{Wenming Cheng}
\credit{Writing -- review \& editing, Funding acquisition}

\affiliation[4]{organization={State Key Laboratory of Bridge Intelligent and Green Construction, Southwest Jiaotong University},
city={Chengdu},
citysep={},
postcode={611756},
country={China}}

\author[1]{Min Xie}
\credit{Writing -- review \& editing}

\author[5]{Xiong Rao}
\credit{Writing -- review \& editing}
\affiliation[5]{organization={School of Mechanical and Electrical Engineering, Guang'an Institute of Technology},
city={Guang'an},
citysep={},
postcode={638000},
country={China}}

\cortext[cor1]{Corresponding author}

\begin{abstract}
Linearized buckling loads of stiffened panels can diverge across commercial shell formulations
because six-degree-of-freedom elements treat drilling rotations and director kinematics differently in the stress stiffness operator.
We derive three analytical stress stiffness operators (membrane-only, uncoupled rotational-block, and shear-coupled full second-variation forms), verified against an exact second-difference energy identity.
Using direct matrix extraction in ANSYS (SHELL181, SHELL281) and modal and strain-energy benchmarks in Abaqus (S4, S8R), we classify these commercial elements across twelve configurations.
On the finest cantilevered shear-panel mesh (in-plane bending and shear), the SHELL181 perturbation and classic procedures predict load factors $11.3\%$ and $28.5\%$ above an independent 3D continuum reference.
By contrast, Abaqus S4 and ANSYS SHELL281 lie $5.2\%$ and $5.8\%$ below this reference and remain indistinguishable from the analytical full form within the junction-idealization uncertainty, while Abaqus S8R lies $20.1\%$ below the reference.
The tested cases indicate that substantial divergence occurs only when non-coplanar junctions carry pre-stress bending or transverse shear and the buckling mode involves junction drilling rotations.
Under pure axial compression or uniform membrane shear flow, the analytical operators coincide and within-element procedures agree within $0.06\%$.
Because buckling under in-plane bending and shear is formulation-dependent, stability design and layout optimization of built-up panels require cross-verification across shell elements or continuum models.
\end{abstract}


\maketitle

\section{Introduction}
\label{sec:intro}

Stiffened shells serve as primary load-bearing structures in aircraft, ships, pressure vessels, cranes, and rail vehicles.
When the skin is thin, structural failure is governed by elastic stability rather than material strength.
Because stiffeners subdivide the skin and control where buckling initiates, structural design depends critically on the underlying buckling solver.
In practice, almost all buckling assessments and layout optimization algorithms solve a linearized eigenvalue problem \citep{ZhouZitong2025a,DongXiaohu2020a,ChuSheng2021a,Gamache2021a}.
This eigenproblem pairs the linear elastic stiffness with a stress stiffness matrix assembled from a static pre-stress state to identify the critical load factor.

Unlike in three-dimensional continuum mechanics, the stress stiffness matrix of a built-up shell structure is not uniquely fixed.
A 3D continuum element has only three translational degrees of freedom (DOFs) per node, so its stress stiffness follows uniquely from the second variation of the strain measure.
In contrast, a flat shell element used for built-up sections requires six engineering DOFs per node, including an in-plane drilling rotation.
At orthogonal stiffener-to-skin junctions, this drilling rotation cannot be omitted because the drilling rotation of one plate coincides with an out-of-plane bending rotation of the adjacent plate.
If the drilling rotation enters the nonlinear shell director, the pre-stress bending moments and transverse shear forces couple the drilling and bending rotations in the stress stiffness operator.
Otherwise, that rotational block vanishes.

Foundational formulations for shell drilling rotations were established by Allman \citep{Allman1984a} and placed on a variational footing by Hughes and Brezzi \citep{Hughes1989a}, with subsequent element families reviewed by Boutagouga \citep{Boutagouga2020a}.
However, structural optimization studies rarely specify the explicit form of their stress stiffness operator.
Commercial finite element theory manuals \citep{AnsysTheory2024,AbaqusTheory2024} likewise document general shell kinematics without stating the exact algebraic structure of the geometric stiffness for elements with drilling rotations.
For prismatic stiffened panels, semi-analytical benchmarks can be established using the exact strip method \citep{Wittrick1974a} or the finite strip method \citep{Cheung1976a}.
For complex built-up topologies featuring dense orthogonal intersections, full shell finite element models and 3D continuum references become indispensable.
Moreover, standard flat-plate verification tests cannot expose differences in rotational stress stiffness because a flat plate under uniform membrane compression has neither bending moments nor transverse shear forces.

Relying on agreement with a single commercial solver is equally insufficient.
Different element formulations within the same software package can assemble distinct stress stiffness operators.
Even within the same element, a classic linear eigenvalue buckling procedure and a linearized perturbation procedure from a nonlinear pre-buckling state need not share the same geometric stiffness matrix.
Matching one commercial load factor merely confirms that an implementation reproduces that specific operator choice.
The essential questions for stiffened-panel analysis are therefore: how far do element formulations and solution procedures separate linearized buckling loads, which mathematical terms account for the discrepancy, and under what physical conditions do these differences become significant?

To address these questions, this study makes three main contributions:
\begin{enumerate}
  \item[(1)] \textbf{Analytical stress stiffness operators:} We derive three geometric stiffness truncations for a six-DOF flat shell element and verify them via an exact second-difference energy identity.
  \item[(2)] \textbf{Commercial operator classification:} Using matrix extraction in ANSYS (SHELL181, SHELL281) and modal/energy benchmarking in Abaqus (S4, S8R), we identify the implicit operators assembled under classic linear and perturbation analyses.
  \item[(3)] \textbf{Continuum benchmarking and activation criteria:} Against 20-node hexahedral continuum baselines across twelve structural configurations, we establish the physical conditions under which shell buckling formulations diverge.
\end{enumerate}

The remainder of this paper is organized as follows.
Section~\ref{sec:formulation} presents the shell element kinematics, the three stress stiffness operators, spectral branch rules, and design sensitivities.
Section~\ref{sec:panels} defines the benchmark structures, model families, and finite element discretizations.
Section~\ref{sec:verif} verifies the analytical operators via the exact energy identity and presents cross-code baseline checks.
Section~\ref{sec:results} presents the operator classifications, 3D continuum benchmarks, and physical activation conditions, followed by the Discussion (Section~\ref{sec:discussion}), Conclusions (Section~\ref{sec:conclusions}), and reproducibility details in Appendices~\ref{app:discrete}--\ref{app:repro}.

\section{Stress stiffness of a six-degree-of-freedom shell}
\label{sec:formulation}

\subsection{Shell element formulation and variants}
\label{sec:femodel}

Both the panel skin and the stiffener walls are discretized with four-node Mindlin--Reissner shell elements that have six engineering DOFs per node.
For an isotropic plate of thickness $t$, plane-stress constitutive matrix $\mat{D}$, and shear modulus $G$, the section rigidities are
\[
  \text{membrane: } t\,\mat{D},
  \qquad
  \text{bending: } \frac{t^{3}}{12}\,\mat{D},
  \qquad
  \text{transverse shear: } k_{\mathrm{s}} G t\,\mat{I}_{2},
\]
where $k_{\mathrm{s}} = 5/6$ is the shear correction factor and $\mat{I}_{2}$ is the $2 \times 2$ identity matrix.
Membrane and bending stiffness terms are integrated using $2\times 2$ Gauss quadrature.
Transverse shear strains are interpolated from the four element edge midpoints using the assumed-strain MITC4 field \citep{Bathe1985a} and integrated by the same $2\times 2$ rule.

In the baseline formulation, the in-plane drilling rotation is regularized by a weak diagonal spring scaled with the bending rigidity, adding $10^{-4}\,\tr[(t^{3}/12)\mat{D}]$ per element to each node's drilling DOF.
Scaling this spring with the much larger membrane rigidity would be admissible for a flat plate.
At orthogonal stiffener junctions, however, rib drilling coincides with skin bending, so a membrane-scaled spring would absorb global bending energy and artificially inflate buckling loads.

Alongside this baseline element, two kinematic variants are defined in Table~\ref{tab:variants} for controlled sensitivity and cross-code checks.
Variant~1 replaces the MITC4 shear field with one-point central reduced integration as a reduced-shear control (Section~\ref{sec:panels}).
Variant~2 enriches the membrane field with condensed incompatible modes~\citep{Taylor1976} to relieve in-plane shear locking and replaces the grounded spring with the Hughes--Brezzi variational drilling penalty~\citep{Hughes1989a}.
Together, these enhancements match the linear kinematics of enhanced commercial four-node shells so that later cross-code checks isolate stress stiffness operator differences (Section~\ref{sec:verif:cross}).

\begin{table}[pos=tbp]
  \centering
  \caption{Shell element variants implemented in the in-house solver.}
  \label{tab:variants}
  \small
  \begin{tabular}{l>{\raggedright\arraybackslash}p{2.5cm}>{\raggedright\arraybackslash}p{2.3cm}>{\raggedright\arraybackslash}p{2.7cm}>{\raggedright\arraybackslash}p{4.1cm}}
    \toprule
    Variant & Membrane field & Transverse shear & Drilling regularization & Primary role \\
    \midrule
    Baseline & Standard bilinear & MITC4 4-point tying & Grounded spring & Analytical reference (Section~\ref{sec:verif:identity}) \\
    Variant 1 & Standard bilinear & 1-point reduced & Grounded spring & Reduced-shear control (Section~\ref{sec:panels}) \\
    Variant 2 & Incompatible modes & MITC4 4-point tying & Hughes--Brezzi penalty & Commercial 4-node shell match (Section~\ref{sec:verif:cross}) \\
    \bottomrule
  \end{tabular}
\end{table}

\subsection{Linearized buckling formulation and the rotational block}
\label{sec:buckling}

Given a static equilibrium displacement $\vect{u}$ under the reference load $\vect{f}$ satisfying $\mat{K}\vect{u} = \vect{f}$, the linearized buckling eigenvalue problem reads
\begin{equation}
  \left[\mat{K} + \lambda\,\Kg(\vect{u})\right]\vect{\varphi} = \vect{0} ,
  \label{eq:eigproblem}
\end{equation}
where $\mat{K}$ is the linear stiffness matrix, $\Kg(\vect{u})$ is the assembled stress stiffness matrix, $\lambda$ is the buckling load factor, and $\vect{\varphi}$ is the buckling mode.
We solve this eigenproblem in the inverse form $\Kg\vect{\varphi} = -\mu\,\mat{K}\vect{\varphi}$ with $\mu = 1/\lambda$, so that critical modes correspond to the eigenvalues of largest magnitude $|\mu|$.

Assembling the stress stiffness matrix for a stiffened panel requires more care than for an isolated flat plate.
Because stiffener walls stand normal to the skin, out-of-plane buckling displacements of a rib are in-plane displacements for the skin, and vice versa.
An operator built solely from transverse displacement gradients would omit the in-plane destabilizing terms governing local wall buckling.
We therefore retain all three translational displacement components at each node.
Let $\mat{N} = [N_{xx}, N_{yy}, N_{xy}]\T$ denote the membrane forces sampled at Gauss point $q$ from the equilibrium state, $g^x_a$ and $g^y_a$ denote the local Cartesian shape function gradients at node $a$, and $w_q$ denote the quadrature weight including the Jacobian determinant.
In the local element frame, the element geometric stiffness takes the Kronecker product form
\begin{equation}
  \Kg^{e} = \mat{M}^{e} \otimes \mathrm{diag}\!\left(\mat{I}_3, \mat{0}_3\right),
  \qquad
  M^{e}_{ab} = \sum_{q} \left[
      N_{xx} g^x_a g^x_b
    + N_{yy} g^y_a g^y_b
    + N_{xy}\left(g^x_a g^y_b + g^y_a g^x_b\right)
  \right] w_q .
  \label{eq:kge}
\end{equation}
Here, $\mat{I}_3$ applies the identical scalar coupling $M^{e}_{ab}$ across all three translational directions, while $\mat{0}_3$ leaves the rotational DOFs uncoupled.
Because $\mat{I}_3$ is invariant under coordinate rotation, Eq.~\eqref{eq:kge} applies uniformly to arbitrarily oriented skin and stiffener elements and reduces to the standard plate expression on flat membranes.

Equation~\eqref{eq:kge} represents only one truncation of the stress stiffness operator, termed the \emph{membrane form}, which is standard for flat shells superposing membrane and plate-bending formulations \citep{Gal2006a}.
However, each shell node has six DOFs (Fig.~\ref{fig:freedoms}): three translations, two bending rotations that tilt the shell director, and an in-plane drilling rotation $\theta_z$ that leaves the director unchanged at first order.
Because an orthogonal rib--skin junction couples one plate's drilling DOF to the adjacent plate's bending rotation (Fig.~\ref{fig:freedoms}b), any second-order director dependence on $\theta_z$ introduces additional rotational terms into $\Kg$.
\begin{figure}[tbp]
  \centering\color{black}
  \begin{tikzpicture}[scale=1.35,
      plate/.style={fill=black!6, draw=black, line width=0.5pt},
      wall/.style={fill=black!14, fill opacity=0.85, draw=black, draw opacity=1, line width=0.5pt},
      ax/.style={-{Stealth[length=4pt]}, draw=black, line width=0.6pt},
      rot/.style={-{Stealth[length=3.5pt]}, draw=black, line width=0.7pt},
      drill/.style={-{Stealth[length=3.5pt]}, draw=black, line width=0.7pt, densely dashed},
      lbl/.style={font=\footnotesize, text=black}]
    \begin{scope}
      \fill[plate] (0,0) -- (2.6,0) -- (3.4,0.9) -- (0.8,0.9) -- cycle;
      \coordinate (P) at (1.7,0.45);
      \draw[ax] (P) -- ++(1.2,0) node[lbl, right] {$x$};
      \draw[ax] (P) -- ++(0.8,0.85) node[lbl, above right=-2pt] {$y$};
      \draw[ax] (P) -- ++(0,1.6) node[lbl, above left=-2pt] {$\mat{e}_3$, the normal};
      \draw[ax, densely dotted] (P) -- ++(0.55,1.45) node[lbl, right=-1pt] {$\mat{d}$, the director};
      \fill (P) circle (1.3pt);
      \draw[rot] ([shift={(0.62,0.55)}]P) arc (0:70:0.62);
      \node[lbl] at ([shift={(0.82,0.62)}]P) {$\theta_y$};
      \draw[rot] ([shift={(-0.42,0.62)}]P) arc (180:100:0.42 and 0.62) node[lbl, left=3pt, yshift=6pt] {$\theta_x$};
      \draw[drill] ([shift={(0.45,-0.03)}]P) arc (0:310:0.45 and 0.17);
      \node[lbl, anchor=north west] at ([shift={(-0.2,-0.15)}]P) {$\theta_z$, drilling};
      \node[lbl, anchor=north] at (1.7,-0.55) {(a) a node of one plate};
    \end{scope}
    \begin{scope}[xshift=5.6cm]
      \fill[plate] (0,0) -- (2.6,0) -- (3.4,0.9) -- (0.8,0.9) -- cycle;
      \coordinate (J) at (1.7,0.35);
      \draw[ax] (J) -- ++(0.85,0.9) node[lbl, above right=-2pt] {$y$};
      \fill[wall] (0.5,0.35) -- (2.9,0.35) -- (2.9,1.75) -- (0.5,1.75) -- cycle;
      \draw[black, line width=0.6pt] (0.5,0.35) -- (2.9,0.35);
      \draw[ax] (J) -- ++(1.1,0) node[lbl, right] {$x$};
      \draw[ax] (J) -- ++(0,1.15) node[lbl, above] {$z$};
      \fill (J) circle (1.3pt);
      \draw[drill] ([shift={(0.45,-0.03)}]J) arc (0:310:0.45 and 0.17);
      \node[lbl] at ([shift={(-1.,1.15)}]J) {rib};
      \node[lbl] at ([shift={(-1.32,-0.2)}]J) {skin};
      \node[lbl] at ([shift={(0.62,-0.2)}]J) {$\theta_z$};
      \draw[drill] ([shift={(0.5,0.62)}]J) arc (0:300:0.5);
      \node[lbl] at ([shift={(0.55,0.92)}]J) {$\theta_y$};
      \draw[rot] ([shift={(-0.2,0.3)}]J) arc (200:-70:0.2 and 0.5);
      \node[lbl, anchor=east] at ([shift={(0.5,0.42)}]J) {$\theta_x$};
      \node[lbl, anchor=west] at (3.05,1.55) {$\theta_x$: bending for both};
      \node[lbl, anchor=west] at (3.05,1.15) {$\theta_y$: bending for the skin,};
      \node[lbl, anchor=west] at (3.35,0.85) {drilling for the wall};
      \node[lbl, anchor=west] at (3.05,0.35) {$\theta_z$: drilling for the skin,};
      \node[lbl, anchor=west] at (3.35,0.05) {bending for the wall};
      \node[lbl, anchor=north] at (1.7,-0.55) {(b) a node where a rib wall meets the skin};
    \end{scope}
  \end{tikzpicture}
  \caption{Nodal rotations of a six-DOF shell element (solid: bending; dashed: drilling): (a)~single plate showing the reference normal $\mat{e}_3$, tilted director $\mat{d}$, and drilling rotation; (b)~orthogonal rib--skin junction sharing $\theta_x$ as a common bending rotation.}
  \label{fig:freedoms}
\end{figure}

In the local element frame $(\mat{e}_1,\mat{e}_2,\mat{e}_3)$, rotating the reference normal $\mat{e}_3$ through the exponential map of the nodal rotation vector $\vect{\theta} = (\theta_x, \theta_y, \theta_z)$ yields the second-order director expansion
\begin{equation}
  \mat{d}^{(2)} = \tfrac{1}{2}\,\vect{\theta}\times(\vect{\theta}\times\mat{e}_3)
  = \tfrac{1}{2}\left(\theta_z\theta_x,\; \theta_z\theta_y,\;
    -\theta_x^2 - \theta_y^2\right) .
  \label{eq:director2}
\end{equation}
Its in-plane components enter both the curvature tensor $\kappa_{\alpha\beta} = \tfrac{1}{2}(\mat{a}_\alpha \cdot \mat{d}_{,\beta} + \mat{a}_\beta \cdot \mat{d}_{,\alpha})$ and the transverse shear strain $\gamma_\alpha = \mat{a}_\alpha \cdot \mat{d}$ ($\alpha,\beta \in \{1,2\}$), where $\mat{a}_1$ and $\mat{a}_2$ are the covariant base vectors of the deformed mid-surface.
Consequently, both strain measures acquire quadratic products pairing the drilling rotation $\theta_z$ with each bending rotation $\theta_\alpha$.
Contracting these second-order strain terms with the frozen pre-stress bending moments $M_{\alpha\beta}$ and transverse shear forces $Q_\alpha$ produces a symmetric rotation-rotation block whose quadratic form $\delta^2 W_{\mathrm{rot}} = \vect{\varphi}\T\mat{K}_{\mathrm{g},\theta\theta}\vect{\varphi}$ (twice the quadratic work, canceling the factor of $1/2$ in Eq.~\eqref{eq:director2}) reads
\begin{equation}
  \delta^2 W_{\mathrm{rot}}
  = \int_A \Big[
      M_{xx}\,(\theta_z\theta_x)_{,x} + M_{yy}\,(\theta_z\theta_y)_{,y}
    + M_{xy}\big((\theta_z\theta_x)_{,y} + (\theta_z\theta_y)_{,x}\big)
    + Q_x\,\theta_z\theta_x + Q_y\,\theta_z\theta_y
    \Big]\,\mathrm{d}A ,
  \label{eq:block}
\end{equation}
integrated over the mid-surface area $A$.
Unlike continuum-derived displacement-gradient approximations that weight rotations by $N_{\alpha\beta} t^2/12$, Eq.~\eqref{eq:block} weights the rotational block strictly by internal bending moments and transverse shear resultants.
We refer to the combination of Eq.~\eqref{eq:kge} and Eq.~\eqref{eq:block} as the \emph{block form}.

Multiplying linear mid-surface displacement gradients with the first-order director increment $\vect{\theta}\times\mat{e}_3$ generates two translation-rotation coupling terms:
\begin{equation}
  \delta^2 W_{\mathrm{tr}}
  = 2\int_A \Big[
      M_{\alpha\beta}\,\mat{v}_{,\alpha}\cdot(\vect{\theta}\times\mat{e}_3)_{,\beta}
    + Q_{\alpha}\,\mat{v}_{,\alpha}\cdot(\vect{\theta}\times\mat{e}_3)
    \Big]\,\mathrm{d}A ,
  \qquad
  \vect{\theta}\times\mat{e}_3 = (\theta_y, -\theta_x, 0) ,
  \label{eq:couplings}
\end{equation}
where $\mat{v}$ denotes the local mid-surface translation vector (with $\alpha,\beta \in \{1,2\}$ summed over in-plane directions).
The leading factor of $2$ appears because the product of a linear base-vector increment with a linear director increment lacks the $1/2$ factor present in Eq.~\eqref{eq:director2}.
Adding both couplings of Eq.~\eqref{eq:couplings} to the membrane and rotational blocks defines the \emph{full form}, representing the complete second variation of pre-stress work for the exponential-map director and the assumed shear field (verified in Section~\ref{sec:verif:identity}).

Together, Eqs.~\eqref{eq:kge}, \eqref{eq:block}, and~\eqref{eq:couplings} define a nested three-level hierarchy of analytical stress stiffness operators (the \emph{membrane}, \emph{block}, and \emph{full} forms).
Independently of this operator hierarchy, the commercial buckling comparisons in Sections~\ref{sec:verif}--\ref{sec:results} distinguish two solver workflows: the \emph{classic linear eigenvalue procedure} (linear buckling about a linear static pre-stress state) and the \emph{linearized perturbation procedure} (eigenvalue extraction about a geometrically nonlinear pre-buckling state).

Two theoretical properties govern the interpretation of Eqs.~\eqref{eq:block} and~\eqref{eq:couplings}.
First, the coefficient $\alpha_{\mathrm{rot}}$ scaling Eq.~\eqref{eq:block} depends on the second-order parameterization of finite rotations: composing drilling before bending leaves $\mat{e}_3$ independent of $\theta_z$ ($\alpha_{\mathrm{rot}} = 0$), reversing that sequence gives $\alpha_{\mathrm{rot}} = 1$, and the exponential map in Eq.~\eqref{eq:director2} gives $\alpha_{\mathrm{rot}} = 1/2$.
All three parameterizations agree to first order and differ only by quadratic coordinate redefinitions ($\vect{\theta} \mapsto \vect{\theta} + \vect{q}(\vect{\theta})$ with terms proportional to $\tfrac{1}{2}\theta_z\theta_\alpha$).
At a true nonlinear bifurcation point, the first variation of total potential energy vanishes and quadratic redefinitions leave the bifurcation load invariant.
In linearized buckling about an unstressed reference state, however, the first variation of pre-stress work is non-zero whenever bending moments or transverse shear forces are present, so varying $\alpha_{\mathrm{rot}}$ directly alters the predicted buckling eigenvalue (quantified on the shear benchmark in Section~\ref{sec:res-continuum}).
This non-uniqueness is intrinsic to linearizing finite rotations about an unstressed reference configuration; consequently, no shell truncation can be judged uniquely ``correct'' in isolation, and its validity must be assessed against internal tangent-stiffness consistency and independent 3D continuum benchmarks.
Unless otherwise stated, all baseline analytical results and benchmarks use the exponential-map value $\alpha_{\mathrm{rot}} = 1/2$, which underlies the exact energy identity in Section~\ref{sec:verif:identity}; Section~\ref{sec:res-continuum} varies this parameter in a controlled sensitivity study.

Parameterization dependence of rotational geometric stiffness is well established in spatial beam formulations \citep{Argyris1979,Argyris1982,YangYeongBin1986} and corotational shells \citep{Rankin1986,NourOmid1991,Haugen1994phd,Felippa2005a}, where it appears through the moment-correction matrix.
The leading term of that correction at the reference configuration is $-\tfrac{1}{2}\mathrm{spin}(\vect{m})$, where $\vect{m} = [m_x, m_y, m_z]\T$ represents the internal nodal moment vector and $\mathrm{spin}(\vect{m})$ is the standard skew-symmetric cross-product matrix.
Although antisymmetric and written without transverse shear terms, it is the closest published analogue of Eq.~\eqref{eq:block}.
Across shell theories, geometrically exact formulations embedding drilling in the 3D rotation tensor \citep{Ibrahimbegovic1994a,Ibrahimbegovic1994b} or as a micropolar DOF \citep{Merlini2011a} retain a resultant-weighted rotational block, whereas variational drilling treatments \citep{Hughes1989a,ZhangTeng2021a} and simplified thin-walled formulations \citep{Senjanovic2012a} omit it.
Equation~\eqref{eq:block} is the facet-element counterpart of the former class under frozen resultants.

Second, regarding assumed-strain transverse shear, the coupling in Eq.~\eqref{eq:couplings} is formulated on the shear field tied at the four element edge midpoints, following the tying strategy of the nonlinear MITC4 shell element \citep{Dvorkin1984a}.
Applying this edge-midpoint tying consistently to both the linear transverse shear stiffness and the second-order stress stiffness variations ensures variational consistency with our adopted assumed-strain energy (Section~\ref{sec:verif:identity}).
While we do not assert complete algebraic equivalence with all nonlinear MITC4 extensions along arbitrary finite-rotation paths, Section~\ref{sec:results} demonstrates that this shear-weighted coupling accounts for the separation between the block and full forms and closely tracks the tested commercial MITC4 implementations.

\subsection{Spectral branches and design sensitivity}
\label{sec:coverage}
\label{sec:sensitivity}

Equation~\eqref{eq:eigproblem} admits eigenvalues of both signs, where negative $\lambda$ denotes buckling under the reversed load direction.
Standard buckling eigensolvers extract either the eigenvalues of largest magnitude $|\mu|$ or only the algebraically largest positive values.
For pure axial compression, extracting the positive branch alone is valid because the reversed load places the panel in tension.
For shear loading, however, both load directions are physically admissible, and on symmetric panels the eigenvalue spectrum is symmetric about zero.

Accordingly, all shear and combined load cases in this study are solved for the dominant modes of largest $|\mu|$ across both branches.
Cross-code comparisons match load factors by magnitude on the active critical branch, and the spectral sign pattern serves as an additional cross-verification metric.
Three evaluation rules follow from this branch symmetry.
First, each reported critical load factor equals the smallest magnitude $|\lambda|$ across both forward and reversed branches.
Second, in perturbation analyses at base-load factor $\alpha$ with incremental load multiplier $\beta$ (Section~\ref{sec:verif:tangent}), the signed total load factor is $\alpha + \beta$, whereas the reported critical load-factor magnitude is $\lambda_{\mathrm{cr}} = |\alpha + \beta|$ (which equals $|\beta| - \alpha$ on a reversed branch with $\beta < -\alpha < 0$).
Third, all modal assurance comparisons pair buckling modes from the identical critical branch so that no mode is compared against its reversed-load mirror image.

When the stress stiffness depends on design variables, the eigenvalue sensitivity comprises three contributions: the explicit derivative of $\Kg$, the elastic derivative of $\mat{K}$, and an adjoint term accounting for pre-stress state dependence.
Incorporating the rotational block of Eq.~\eqref{eq:block} introduces an additional adjoint load equal to the derivative of $\vect{\varphi}\T\Kg(\vect{u})\vect{\varphi}$ with respect to the equilibrium displacement $\vect{u}$.
Its moment-weighted contribution acts through the bending curvature operator, whereas its shear-weighted contribution is assembled consistently at the four MITC4 edge-midpoint tying locations.
Appendix~\ref{app:discrete} provides the discrete element matrix of the rotational block, and Appendix~\ref{app:adjoint} derives both adjoint sensitivity terms required for gradient-based optimization.
These sensitivity expressions apply strictly to separated eigenvalues and are verified end-to-end against central finite differences in Appendix~\ref{app:adjoint}.

\section{Benchmark structures}
\label{sec:panels}

All benchmark structures---comprising stiffened rectangular panels (Figs.~\ref{fig:panel}--\ref{fig:designs}), additional thin-walled shells and open beams (Fig.~\ref{fig:newmodels}), and a full-scale box girder (Fig.~\ref{fig:boxgirder})---use isotropic steel ($E_0 = 2.1 \times 10^{11}$~Pa, $\nu = 0.3$).

\subsection{Stiffened panel geometries and loading}
\label{sec:panels:primary}

All stiffened-panel benchmarks share the baseline footprint of Fig.~\ref{fig:panel}; in the cantilevered load cases, the edge $x = 0$ is fully clamped in all six DOFs and a resultant load of $F = 10^{5}$~N is distributed over the skin nodes along $x = L_x$.
Three loading cases on the cantilevered panel isolate distinct pre-stress states: axial compression along $-x$ generates a nearly pure membrane stress field, whereas in-plane end shear along $+y$ induces mean shear $N_{xy} \sim F/L_y$ together with substantial in-plane bending.
Throughout this study, this bending-plus-shear load state is referred to as \emph{in-plane bending and shear} (or \emph{in-plane end shear} when distinguishing load directions), as distinct from pure \emph{membrane shear flow}, which applies self-equilibrated tangential edge tractions on all four sides without in-plane bending (Section~\ref{sec:res-when}).
A combined loading case applies both reference loads simultaneously along $-x$ and $+y$ to provide intermediate stress gradations; reversing the combined-case load vector across the two spectral branches (Section~\ref{sec:coverage}) reverses both compression and shear simultaneously rather than reversing shear under fixed compression.

Under the reference load, axial compression remains elastic up to buckling at $\lambda \approx 11$ ($\sim 64$~MPa, well below Q235/S235 yield at $235$~MPa), whereas in-plane end shear reaches Q235/S235 yield at the clamped root around $\lambda \approx 2.4$ (and Q355/S355 at roughly half the elastic buckling load).
Because linearized buckling ratios are independent of load magnitude, the reported shear and combined load factors serve strictly as elastic operator benchmarks.

\begin{figure}[tbp]
  \centering\color{black}
  \resizebox{\linewidth}{!}{%
  \begin{tikzpicture}[scale=1.0,
      skin/.style={fill=black!6, draw=black, line width=0.5pt},
      grid/.style={draw=black!25, line width=0.2pt},
      rib/.style={draw=black, line width=1.1pt},
      clamp/.style={pattern=north east lines, pattern color=black!60},
      ld/.style={-{Stealth[length=4.5pt]}, draw=black, line width=0.8pt},
      dim/.style={<->, >={Stealth[length=3.5pt]}, draw=black, line width=0.35pt},
      lbl/.style={font=\footnotesize, text=black}]
    \def\PW{4.2} \def\PH{1.4}
    \foreach \cx/\name/\case in {0/a/axial, 5.4/b/shear, 10.8/c/combined}{
      \begin{scope}[xshift=\cx cm]
        \fill[skin] (0,0) rectangle (\PW,-\PH);
        \foreach \i in {1,...,23}{\draw[grid] ({\i*\PW/24},0) -- ({\i*\PW/24},-\PH);}
        \foreach \j in {1,...,7}{\draw[grid] (0,{-\j*\PH/8}) -- (\PW,{-\j*\PH/8});}
        \draw[black, line width=0.5pt] (0,0) rectangle (\PW,-\PH);
        \fill[clamp] (-0.18,0.08) rectangle (0,-\PH-0.08);
        \draw[black, line width=0.7pt] (0,0.08) -- (0,-\PH-0.08);
        \node[lbl, anchor=north] at ({\PW/2},{-\PH-0.12}) {(\name) \case};
      \end{scope}
    }
    \foreach \j in {0,...,4}{\draw[ld] ({\PW+0.55},{-\j*\PH/4}) -- ({\PW+0.08},{-\j*\PH/4});}\node[lbl, anchor=south] at ({\PW+0.42},0.16) {\scriptsize $10^{5}$~N};
    \foreach \j in {0,...,4}{\draw[ld] ({5.4+\PW+0.14},{-\j*\PH/4+0.16}) -- ({5.4+\PW+0.14},{-\j*\PH/4-0.14});}\node[lbl, anchor=south] at ({5.4+\PW+0.14},0.18) {\scriptsize $10^{5}$~N};
    \foreach \j in {0,...,4}{
      \draw[ld] ({10.8+\PW+0.55},{-\j*\PH/4}) -- ({10.8+\PW+0.08},{-\j*\PH/4});
      \draw[ld] ({10.8+\PW+0.72},{-\j*\PH/4+0.16}) -- ({10.8+\PW+0.72},{-\j*\PH/4-0.14});}\node[lbl, anchor=south] at ({10.8+\PW+0.42},0.18) {\scriptsize $10^{5}$~N each};
    \draw[dim] (0,0.32) -- (\PW,0.32) node[midway, above=-1pt, lbl] {$L_x = 3$~m};
    \draw[dim] (-0.42,0) -- (-0.42,-\PH) node[midway, left=-1pt, lbl] {$L_y = 1$~m};
    \node[lbl, anchor=west] at (0.05,0.12) {\scriptsize $x$ \,$\rightarrow$};
    \node[lbl, anchor=west] at (-0.05,-0.35) {\scriptsize $y\downarrow$};
    \begin{scope}[xshift=17.15cm, yshift=-0.05cm]
      \fill[black!6, draw=black, line width=0.45pt] (-0.9,-1.25) rectangle (0.9,-1.40);
      \draw[black!60, densely dashed, line width=0.35pt] (-0.9,-1.325) -- (0.9,-1.325);
      \fill[black!18, draw=black, line width=0.45pt] (-0.06,-1.25) rectangle (0.06,0.0);
      \draw[black, line width=0.9pt] (0,-1.325) -- (0,0.0);
      \fill[black] (0,-1.325) circle (0.03);
      \draw[dim] (0.25,-1.325) -- (0.25,0.0) node[midway, right=-1pt, lbl] {\scriptsize $H_z = 0.1$~m};
      \node[lbl, anchor=west] at (0.92,-1.325) {\scriptsize $t = 10$~mm};
      \node[lbl, anchor=south] at (0,0.02) {\scriptsize $t_{\mathrm{s}} = 8$~mm};
      \node[lbl, anchor=north, align=center] at (0,-1.45) {\scriptsize (d) section: wall tied\\[-2pt]\scriptsize to the skin mid-surface};
    \end{scope}
  \end{tikzpicture}}
  \caption{The stiffened panel: (a) axial, (b) shear and (c) combined loading of the edge
  $x = L_x$, and (d) the section at a wall foot. Not to scale across the section.}
  \label{fig:panel}
\end{figure}

\begin{figure}[tbp]
  \centering
  \begin{subfigure}{0.32\linewidth}
    \centering
    \includegraphics[width=\linewidth]{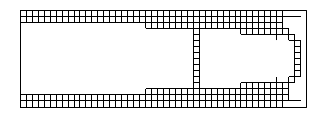}
    \caption{shear, design A (block form)}
    \label{fig:design-A}
  \end{subfigure}\hfill
  \begin{subfigure}{0.32\linewidth}
    \centering
    \includegraphics[width=\linewidth]{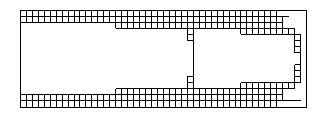}
    \caption{shear, design B (membrane form)}
    \label{fig:design-B}
  \end{subfigure}\hfill
  \begin{subfigure}{0.32\linewidth}
    \centering
    \includegraphics[width=\linewidth]{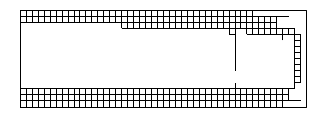}
    \caption{combined case}
    \label{fig:design-C}
  \end{subfigure}\\[1.5ex]
  \begin{subfigure}{0.32\linewidth}
    \centering
    \includegraphics[width=\linewidth]{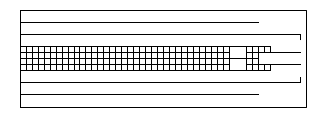}
    \caption{axial compression}
    \label{fig:design-X}
  \end{subfigure}
  \caption{The four discrete designs on which the operators are compared, drawn at one scale on the
  $48 \times 16$ grid. The right edge is loaded, the left edge clamped.}
  \label{fig:designs}
\end{figure}

Four topologically optimized panels (Fig.~\ref{fig:designs}) serve as demanding benchmark structures with dense non-coplanar junctions: one under axial compression, one under combined loading, and two under in-plane bending and shear.
These discrete layouts originate from a separate stiffener layout optimization study, generated on a $48 \times 16$ candidate rib lattice at a stiffener volume fraction of $0.4$ by maximizing the critical buckling load factor under SIMP density penalization ($p = p_G = 3$, $\Emin = 10^{-9}$).
Whereas layout optimization methodology lies outside the scope of this work, the present manuscript treats these four layouts strictly as fixed benchmark geometries (with complete discrete segment tables provided in the Zenodo reproducibility archive) to isolate the behavior of stress stiffness operators.
Shear designs~A and~B differ only in the stress stiffness operator used during layout generation (block form for design~A vs.\ membrane form for design~B), testing whether operator selection alters the resulting rib layout as well as reported load factors.

\subsection{Structural model families}
\label{sec:panels:models}

Three structural model families are evaluated across these panels (Table~\ref{tab:models}):
\begin{enumerate}
  \item \textbf{Tied optimization models:} Used during stiffener layout generation and controlled tied-grid operator studies in Section~\ref{sec:results}. Here, candidate stiffener walls on the background grid are tied to the skin mid-surface via multipoint constraints on all six degrees of freedom:
  \begin{equation}
    \vect{u}_d = (1-\xi)\,\vect{u}_1 + \xi\,\vect{u}_2 .
    \label{eq:mpc}
  \end{equation}
  In this relation, $\vect{u}_d$ represents the six freedoms of a dependent wall foot node, $\vect{u}_1$ and $\vect{u}_2$ are the bounding skin nodes, and $\xi \in [0,1]$ is its normalized position along the skin edge. In the in-house solver, candidate rib segments below the $0.5$ density threshold are retained with a soft material penalty ($\Emin$ in $\mat{K}$ and SIMP scaling in $\Kg$); in exported commercial decks (ANSYS and Abaqus), all void elements are strictly omitted prior to export.
  \item \textbf{Shared-node rebuilt models:} Used for all primary cross-code classifications, mesh convergence studies, and primary commercial benchmarks in ANSYS and Abaqus. The optimized layout is converted into explicit CAD boundary lines and remeshed as discrete shell mid-surfaces sharing common nodes along all intersection lines. Void and intermediate-density elements are \emph{strictly and entirely excluded}, eliminating multipoint constraints and soft-element regularizations.
  \item \textbf{Three-dimensional continuum models:} Independent benchmark models constructed in 20-node hexahedral elements (SOLID186 in ANSYS and C3D20 in Abaqus). These models discretize the physical three-dimensional volume of the skin and standing rib walls with shared nodes and no rotational degrees of freedom, free of shell director and drilling-rotation assumptions and void elements, while retaining an explicit geometric idealization of the rib--skin junction.
\end{enumerate}

\begin{table}[pos=!htbp]
  \centering
  \caption{Model families evaluated in this study, their void-element treatment, junction idealization, and formulations.}
  \label{tab:models}
  \small
  \begin{tabular}{llll}
    \toprule
    Model family & Void elements & Rib--skin junction & Formulations \\
    \midrule
    Tied optimization & $\Emin$ (in-house); none (export) & MPC constraints & In-house (all), SHELL181, S4/S4R \\
    Shared-node rebuild & Strictly excluded & Shared nodes & In-house, SHELL181/281, S4, S8R \\
    3D continuum & Strictly excluded & Conforming solid & SOLID186, C3D20 \\
    \bottomrule
  \end{tabular}
\end{table}

In the tied optimization models, exported commercial decks enforce Eq.~\eqref{eq:mpc} via linear constraint equations (Appendix~\ref{app:repro}). The in-house solver enforces Eq.~\eqref{eq:mpc} via penalty stiffness $k_{\mathrm{mpc}} = \alpha_f \Kref$ ($\alpha_f = 10^3$, with $\Kref$ the maximum skin-node diagonal stiffness) and applies a weak rotational regularization exclusively to rotational DOFs to prevent drilling ill-conditioning:
\begin{equation}
  \mat{K} \leftarrow \mat{K} + \varepsilon_r \Kref \sum_{i \in \mathcal{R}} \mat{e}_i \mat{e}_i\T ,
  \qquad \varepsilon_r = 10^{-11}.
  \label{eq:rotreg}
\end{equation}
In all shell models, the stiffener wall connects directly to the skin mid-surface without modeling the 5~mm half-skin thickness offset (Fig.~\ref{fig:panel}d); Section~\ref{sec:res-continuum} quantifies the geometric sensitivity of this idealization.

\subsection{Additional thin-walled benchmarks and full-scale box girder}
\label{sec:panels:additional}

Additional benchmark structures enter in two groups to isolate junction density, shell curvature, and cross-code generality (Fig.~\ref{fig:newmodels}).
First, a conventionally stiffened \emph{conforming panel} and three derivatives (cut ribs with free ends, halved rib pitch, and pure membrane shear flow) isolate junction density, while a densely stiffened panel validates pure axial agreement (Table~\ref{tab:fulldensity}).
Multi-level mesh-convergence and matrix-export checks use shared-node rebuilds of the conforming panel, shear design~A, and the axial design.
Second, plain and stiffened cylinders, an I-beam on forks, and a channel cantilever under eccentric tip shear probe configurations lacking dense orthogonal plate intersections.

\begin{figure}[tbp]
  \centering\color{black}
  \resizebox{\linewidth}{!}{%
  \begin{tikzpicture}[scale=1.0,
      shell/.style={fill=black!6, draw=black, line width=0.5pt},
      face/.style={fill=black!14, draw=black, line width=0.5pt},
      hid/.style={draw=black!50, line width=0.3pt, densely dashed},
      clamp/.style={pattern=north east lines, pattern color=black!60},
      ld/.style={-{Stealth[length=4.5pt]}, draw=black, line width=0.8pt},
      dim/.style={<->, >={Stealth[length=3.5pt]}, draw=black, line width=0.35pt},
      lbl/.style={font=\footnotesize, text=black}]
    \begin{scope}
      \def\R{0.95} \def\L{2.6} \def\ex{0.26}
      \fill[shell] (0,-\R) -- (\L,-\R) arc[start angle=-90, end angle=90, x radius=\ex, y radius=\R]
        -- (0,\R) arc[start angle=90, end angle=270, x radius=\ex, y radius=\R] -- cycle;
      \foreach \xr in {0.25,0.5,0.75}{
        \draw[hid] ({\xr*\L},-\R) arc[start angle=-90, end angle=90, x radius=\ex, y radius=\R];}
      \draw[black, line width=0.5pt] (\L,0) ellipse[x radius=\ex, y radius=\R];
      \fill[clamp] (-0.36,-\R-0.1) rectangle (-0.18,\R+0.1);
      \draw[black, line width=0.7pt] (-0.18,-\R-0.1) -- (-0.18,\R+0.1);
      \draw[black, line width=0.5pt] (0,0) ellipse[x radius=\ex, y radius=\R];
      \foreach \yy in {0.62,-0.62}{\draw[ld] ({\L+0.75},\yy) -- ({\L+0.28},\yy);}
      \draw[ld] ({\L+0.55},{0.25}) arc[start angle=60, end angle=-240, x radius=0.12, y radius=0.30];
      \node[lbl, anchor=west] at ({\L+0.78},0.62) {$P$};
      \node[lbl, anchor=west] at ({\L+0.72},-0.05) {$T$};
      \draw[dim] (0,{\R+0.25}) -- (\L,{\R+0.25}) node[midway, above=-1pt, lbl] {$L = 1$~m};
    \end{scope}
    \begin{scope}[xshift=5.05cm]
      \def\R{0.95} \def\d{0.17}
      \fill[black!14] (0,0) circle (\R);
      \fill[white] (0,0) circle ({\R-\d});
      \draw[black, line width=0.5pt] (0,0) circle ({\R-\d});
      \draw[black, line width=0.9pt] (0,0) circle (\R);
      \foreach \k in {0,...,23}{\draw[black, line width=0.8pt] ({15*\k}:\R) -- ({15*\k}:{\R-\d});}
      \draw[dim] (0,0) -- (-52:{\R-\d}) node[midway, above right=-2pt, lbl] {$R$};
      \node[font=\scriptsize, text=black, anchor=south] at (0,{\R+0.16}) {stiffened section};
    \end{scope}
    \node[lbl, anchor=north] at (2.82,-1.10) {(a) cylinder (plain / stiffened)};
    \begin{scope}[xshift=8.05cm, yshift=-0.55cm]
      \def\px{1.0} \def\qx{0.45} \def\qy{0.36}
      \def\LL{3.0} \def\b{0.50} \def\h{1.10}
      \def\PT#1#2#3{({(#1)*\px+(#2)*\qx},{(#2)*\qy+(#3)})}
      \fill[face] \PT{0}{0}{0} -- \PT{\LL}{0}{0} -- \PT{\LL}{\b}{0} -- \PT{0}{\b}{0} -- cycle;
      \fill[shell] \PT{0}{0}{0} -- \PT{\LL}{0}{0} -- \PT{\LL}{0}{\h} -- \PT{0}{0}{\h} -- cycle;
      \fill[face] \PT{0}{-\b}{0} -- \PT{\LL}{-\b}{0} -- \PT{\LL}{0}{0} -- \PT{0}{0}{0} -- cycle;
      \fill[face] \PT{0}{-\b}{\h} -- \PT{\LL}{-\b}{\h} -- \PT{\LL}{\b}{\h} -- \PT{0}{\b}{\h} -- cycle;
      \foreach \xx in {0,\LL}{
        \draw[black, line width=1.1pt] \PT{\xx}{-\b}{\h} -- \PT{\xx}{\b}{\h};
        \draw[black, line width=1.1pt] \PT{\xx}{-\b}{0} -- \PT{\xx}{\b}{0};
        \draw[black, line width=1.1pt] \PT{\xx}{0}{0} -- \PT{\xx}{0}{\h};
        \draw[black] \PT{\xx}{0}{0} ++(0,-0.02) -- ++(-0.10,-0.17) -- ++(0.20,0) -- cycle;}
      \draw[ld] \PT{\LL+0.30}{0}{\h+0.05} arc[start angle=90, end angle=-90, x radius=0.28, y radius={0.5*\h+0.05}];
      \draw[ld] \PT{-0.30}{0}{\h+0.05} arc[start angle=90, end angle=270, x radius=0.28, y radius={0.5*\h+0.05}];
      \node[lbl, anchor=west] at \PT{\LL+0.62}{0}{0.5*\h} {$M$};
      \node[lbl, anchor=east] at \PT{-0.62}{0}{0.5*\h} {$M$};
      \draw[dim] \PT{0}{\b}{\h+0.22} -- \PT{\LL}{\b}{\h+0.22} node[midway, above=-1pt, lbl] {$L = 4$~m};
      \node[lbl, anchor=north] at ({0.5*\LL},-0.55) {(b) I-beam on forks};
    \end{scope}
    \begin{scope}[xshift=13.2cm, yshift=-0.55cm]
      \def\px{1.0} \def\qx{0.45} \def\qy{0.36}
      \def\LL{2.2} \def\b{0.42} \def\h{1.00}
      \def\PT#1#2#3{({(#1)*\px+(#2)*\qx},{(#2)*\qy+(#3)})}
      \fill[clamp] \PT{-0.18}{0.15}{-0.30} -- \PT{0}{0.15}{-0.30} -- \PT{0}{0.15}{\h+0.12} -- \PT{-0.18}{0.15}{\h+0.12} -- cycle;
      \draw[black, line width=0.7pt] \PT{0}{0.15}{-0.30} -- \PT{0}{0.15}{\h+0.12};
      \fill[shell] \PT{0}{0}{0} -- \PT{\LL}{0}{0} -- \PT{\LL}{0}{\h} -- \PT{0}{0}{\h} -- cycle;
      \fill[face] \PT{0}{-\b}{0} -- \PT{\LL}{-\b}{0} -- \PT{\LL}{0}{0} -- \PT{0}{0}{0} -- cycle;
      \fill[face] \PT{0}{-\b}{\h} -- \PT{\LL}{-\b}{\h} -- \PT{\LL}{0}{\h} -- \PT{0}{0}{\h} -- cycle;
      \draw[black, line width=1.1pt] \PT{\LL}{-\b}{\h} -- \PT{\LL}{0}{\h} -- \PT{\LL}{0}{0} -- \PT{\LL}{-\b}{0};
      \foreach \zz in {0.25,0.60,0.95}{\draw[ld] \PT{\LL+0.12}{0}{\zz} -- \PT{\LL+0.12}{0}{\zz-0.28};}
      \node[lbl, anchor=west] at \PT{\LL+0.16}{0}{0.55} {$F$};
      \draw[dim] \PT{0}{0}{\h+0.30} -- \PT{\LL}{0}{\h+0.30} node[midway, above=-1pt, lbl] {$L = 2$~m};
      \node[lbl, anchor=north] at ({0.5*\LL},-0.55) {(c) channel cantilever};
    \end{scope}
  \end{tikzpicture}
  }
  \caption{Additional thin-walled benchmarks: (a)~cylinder under axial load $P$ or torque $T$ (dashed quarter rings and section apply to the stiffened cylinder),
  (b)~I-beam on forks under end moments, and (c)~channel cantilever under a tip load.}
  \label{fig:newmodels}
\end{figure}

Finally, a full-scale container-crane box girder (Fig.~\ref{fig:boxgirder}) tests rotational block sensitivity in a realistic diaphragm-stiffened engineering girder.

\begin{figure}[tbp]
  \centering\color{black}
  \begin{tikzpicture}[scale=1.0,
      skin/.style={fill=black!6, draw=black, line width=0.5pt},
      face/.style={fill=black!14, draw=black, line width=0.5pt},
      oh/.style={fill=black!10, draw=black, line width=0.5pt},
      dia/.style={draw=black!55, line width=0.3pt, densely dashed},
      dim/.style={<->, >={Stealth[length=4pt]}, draw=black, line width=0.35pt},
      ld/.style={-{Stealth[length=5pt]}, draw=black, line width=0.7pt},
      sup/.style={regular polygon, regular polygon sides=3, draw=black, fill=white,
                  inner sep=2.0pt, anchor=north},
      lbl/.style={font=\footnotesize, text=black}]

    \def\px{1.00} \def\qx{0.42} \def\qy{0.34}
    \def\LL{7.2}  \def\W{0.82}  \def\H{1.50} \def\ov{0.10}
    \def\eo{0.21} \def\bp{0.21}  
    \newcommand{\PT}[3]{({(#1)*\px+(#2)*\qx},{(#2)*\qy+(#3)})}

    \fill[skin] \PT{-\eo}{0}{0} -- \PT{\LL+\eo}{0}{0} -- \PT{\LL+\eo}{0}{\H}
      -- \PT{-\eo}{0}{\H} -- cycle;
    \fill[face] \PT{\LL+\eo}{0}{0} -- \PT{\LL+\eo}{\W}{0} -- \PT{\LL+\eo}{\W}{\H}
      -- \PT{\LL+\eo}{0}{\H} -- cycle;
    \fill[oh] \PT{-\eo}{-\ov}{0} -- \PT{\LL+\eo}{-\ov}{0} -- \PT{\LL+\eo}{0}{0}
      -- \PT{-\eo}{0}{0} -- cycle;
    \fill[face] \PT{-\eo}{-\ov}{\H} -- \PT{\LL+\eo}{-\ov}{\H}
      -- \PT{\LL+\eo}{\W+\ov}{\H} -- \PT{-\eo}{\W+\ov}{\H} -- cycle;
    \draw[black, line width=0.5pt] \PT{-\eo}{0}{\H} -- \PT{\LL+\eo}{0}{\H};

    \foreach \i in {1,...,5}{
      \pgfmathsetmacro{\sx}{\i*\LL/6}
      \draw[dia] \PT{\sx}{0}{0} -- \PT{\sx}{0}{\H} -- \PT{\sx}{\W}{\H};
    }

    \fill[black!30, draw=black, line width=0.35pt] \PT{-\bp}{-\ov}{-0.06} -- \PT{\bp}{-\ov}{-0.06}
      -- \PT{\bp}{-\ov}{0} -- \PT{-\bp}{-\ov}{0} -- cycle;
    \fill[black!30, draw=black, line width=0.35pt] \PT{\LL-\bp}{-\ov}{-0.06} -- \PT{\LL+\bp}{-\ov}{-0.06}
      -- \PT{\LL+\bp}{-\ov}{0} -- \PT{\LL-\bp}{-\ov}{0} -- cycle;
    \foreach \k/\xx in {1/-0.12, 2/0.12}{
      \node[sup] at \PT{\xx}{-\ov}{-0.06} {};
      \node[sup] (SR\k) at \PT{\LL+\xx}{-\ov}{-0.06} {};
      \node[circle, draw=black, inner sep=0.8pt, anchor=north] at (SR\k.south) {};
    }
    \node[lbl, anchor=north east, xshift=6pt, yshift=-10pt] at \PT{-\bp}{-\ov}{-0.06}
      {$2.0$~m pad, axially held};
    \node[lbl, anchor=north west, xshift=-14pt, yshift=-16pt] at \PT{\LL+\bp}{-\ov}{-0.06}
      {$2.0$~m pad, axially free};

    \foreach \dx in {-0.926,0.926}{
      \pgfmathsetmacro{\sx}{0.5*\LL+\dx}
      \fill[black!35, draw=black, line width=0.35pt] \PT{\sx-0.16}{\W-0.14}{\H} -- \PT{\sx+0.16}{\W-0.14}{\H}
        -- \PT{\sx+0.16}{\W+0.14}{\H} -- \PT{\sx-0.16}{\W+0.14}{\H} -- cycle;
      \draw[ld] \PT{\sx}{\W}{\H+0.56} -- \PT{\sx}{\W}{\H+0.06};
    }
    \draw[dim] \PT{0.5*\LL-0.926}{\W}{\H+0.62} -- \PT{0.5*\LL+0.926}{\W}{\H+0.62}
      node[midway, above=1pt, lbl] {$9.0$~m};
    \node[lbl, anchor=south] at \PT{0.5*\LL}{\W}{\H+0.92}
      {$2 \times 200$~kN on two pads, above the $10$~mm web};

    \draw[dim] \PT{0}{-\ov}{-0.55} -- \PT{\LL}{-\ov}{-0.55}
      node[midway, below, lbl] {$35$~m};
    \draw[dim] ({\LL*\px+(\W+\ov)*\qx+0.30},{(\W+\ov)*\qy}) --
               ({\LL*\px+(\W+\ov)*\qx+0.30},{(\W+\ov)*\qy+\H})
      node[midway, right=1pt, lbl] {$2.2$~m};
    \draw[dim] ({\LL*\px+0.85},{-0.12}) -- ({\LL*\px+\W*\qx+0.85},{\W*\qy-0.12})
      node[midway, right=2pt, lbl, xshift=-4pt] {$1.2$~m};

  \end{tikzpicture}
  \caption{The benchmark box girder, to the specification of Section~\ref{sec:panels} and Appendix~\ref{app:repro}; the interior
  diaphragms are traced dashed and their manholes are not drawn. Not to scale along the
  span.}
  \label{fig:boxgirder}
\end{figure}

Table~\ref{tab:structures} compiles the remaining cross-section dimensions, load magnitudes, and discretizations complementing Figs.~\ref{fig:panel}--\ref{fig:boxgirder}, where L1--3 denote shared-node mesh subdivision levels~1--3 (Appendix~\ref{app:repro}).
Specifically, the optimized shear panels feature over $350$ crossing or T-junctions compared with $61$ on the baseline conforming panel; complete node-topology counts and box-girder reconstruction details appear in Appendix~\ref{app:repro}.

\begin{table}[pos=!htbp]
\centering
\caption{Additional dimensions, load magnitudes, and discretizations complementing Figs.~\ref{fig:panel}--\ref{fig:boxgirder}.}
\label{tab:structures}
\footnotesize
\setlength{\tabcolsep}{4pt}
\begin{tabular}{>{\raggedright\arraybackslash}p{2.9cm}>{\raggedright\arraybackslash}p{4.4cm}>{\raggedright\arraybackslash}p{3.3cm}>{\raggedright\arraybackslash}p{4.3cm}}
\toprule
Structure / Case & Additional Geometry & Load Magnitude & Discretization \\
\midrule
Optimized panels: A, B, Combined, Axial (Figs.~\ref{fig:panel}, \ref{fig:designs}) &
Vol.\ frac.\ $0.4$; $641$ (design~A) / $640$ (others) rib segments &
As Fig.~\ref{fig:panel} &
$48\times 16$ lattice ($0.0625$~m pitch; $10\,368$ cand., $4\,614$ active on A); rebuilds L1--3 (A), L1--2 (axial); 3D solid \\[3pt]
Conforming panel \& variants &
$250$~mm pitch ($448$ seg.); cut ribs ($36$ free ends); $125$~mm pitch &
As Fig.~\ref{fig:panel} (shear), or pure edge shear flow &
Tied ($3\,456$ active, Variant~2); shared-node rebuilds L1--2 \\[3pt]
Dense axial panel (Table~\ref{tab:fulldensity}) &
Full rib grid on $24\times 8$ lattice &
As Fig.~\ref{fig:panel} (axial) &
$1$ shell element per grid cell ($24\times 8$) \\[3pt]
Cylinders, plain \& stiffened (Fig.~\ref{fig:newmodels}a) &
$R = 0.5$~m, $t = 5$~mm; stiffened: $24$ stringers + quarter rings ($40\times 4$~mm) &
$P = 5\times 10^{6}$~N or $T = 10^{6}$~N\,m &
$96\times 32$ cells on cylinder wall and stiffeners \\[3pt]
I-beam on forks (Fig.~\ref{fig:newmodels}b) &
$400\times 200$~mm, $12$~mm flanges, $8$~mm web &
$M = 10^{5}$~N\,m &
Uniform $25$~mm shell cells \\[3pt]
Channel cantilever (Fig.~\ref{fig:newmodels}c) &
$300\times 100$~mm, $10$~mm flanges, $6$~mm web &
$F = 2\times 10^{4}$~N &
Uniform $25$~mm shell cells \\[3pt]
Box girder (Fig.~\ref{fig:boxgirder}) &
$1.0$~m overhangs; $14$~mm flanges, $8$/$10$~mm webs; $34$ diaphragms &
Self-weight $261$~kN (plus Fig.~\ref{fig:boxgirder} wheel loads) &
$333\times 24\times 44$ grid ($67\,716$ shell elems; MITC4) \\
\bottomrule
\end{tabular}
\end{table}

\section{Solver verification and cross-code comparison}
\label{sec:verif}

Verification proceeds in two stages.
First, Section~\ref{sec:verif:identity} establishes internal mathematical consistency by checking whether each assembled stress stiffness matrix satisfies the exact second variation of the element's internal work.
Second, Section~\ref{sec:verif:cross} establishes baseline cross-code agreement by confirming that our shell models reproduce commercial solver predictions closely whenever rotational stress stiffness contributions are negligible.
Direct matrix extraction and modal identification of the implicit commercial operators follow in Section~\ref{sec:results}.

\subsection{Exact energy identity for stress stiffness verification}
\label{sec:verif:identity}

In the shell formulation, the linear elastic stiffness is the second derivative of the internal strain energy at the undeformed configuration.
Correspondingly, the consistent stress stiffness must equal the second derivative of the nonlinear internal pre-stress work $W(\vect{\eta})$ with all stress resultants held frozen at their pre-buckling values:
\begin{equation}
W(\vect{\eta}) = \int_A \left( N_{\alpha\beta}\,\varepsilon_{\alpha\beta}(\vect{\eta})
+ M_{\alpha\beta}\,\kappa_{\alpha\beta}(\vect{\eta})
+ Q_{\alpha}\,\gamma_{\alpha}(\vect{\eta}) \right) \mathrm{d}A ,
\qquad
\vect{\varphi}\T \Kg\, \vect{\varphi}
= \left. \frac{\mathrm{d}^{2}}{\mathrm{d}s^{2}} W(s\vect{\varphi}) \right|_{s=0} .
\label{eq:identity}
\end{equation}
Here, $\vect{\eta}$ is a global nodal perturbation vector scaled by scalar $s$, while $N_{\alpha\beta}$, $M_{\alpha\beta}$, and $Q_\alpha$ are the frozen membrane, moment, and transverse shear resultants.
The strain measures $\varepsilon$, $\kappa$, and $\gamma$ correspond to the second-order director truncation of Eq.~\eqref{eq:director2}, rather than linearized strain increments.
Any stress stiffness form satisfying Eq.~\eqref{eq:identity} is variationally consistent with the element's internal kinematics, whereas other forms violate that identity independently of commercial software comparisons.

Two mathematical properties turn Eq.~\eqref{eq:identity} into an exact algebraic check.
First, $W(\vect{\eta})$ is a scalar functional that requires no matrix assembly to evaluate.
Second, under the quadratic director expansion of Eq.~\eqref{eq:director2}, each truncated strain measure is at most cubic in $\vect{\eta}$ (from multiplying linear mid-surface displacement gradients with quadratic director rotations).
Consequently, in the central second difference $[\,W(h\vect{\varphi}) - 2W(\vect{0}) + W(-h\vect{\varphi})\,]/h^{2}$, all odd cubic terms cancel identically for any perturbation step $h$.
There is no trade-off between truncation error and round-off: step sizes $h = 10^{-2}$, $10^{-1}$, and $1$ agree to thirteen significant figures.

To prevent circular validation, the scalar functional $W$ is evaluated independently from the finite element kinematics alone, sharing only the local element coordinate frame with the matrix assembly.
For the transverse shear energy, both the nonlinear shear strain $\gamma_\alpha$ and the frozen shear force $Q_\alpha$ must come from the same MITC4 edge-midpoint tying field.
If $\gamma_\alpha$ is evaluated pointwise instead, the reference disagrees with the assembled rotational block by tens of percent.
Similarly, evaluating the frozen shear force at the element center (a one-point rule) misses the assembled block form by $13\%$ on pure rotational test vectors.
Evaluating both quantities consistently from the assumed-strain field closes the identity to machine precision, as verified in Table~\ref{tab:identity}.

Table~\ref{tab:identity} evaluates Eq.~\eqref{eq:identity} across prescribed global test vectors on a small cantilevered stiffened panel ($0.25$~m square, ribs on a $4 \times 4$ interior grid) under in-plane end shear.
Using global test components ensures that an out-of-plane bending rotation on the skin acts simultaneously as a drilling rotation on the orthogonal ribs.
For comparison, the \emph{continuum truncation} column (distinct from the 3D solid model of Section~\ref{sec:res-continuum}) evaluates Eq.~\eqref{eq:kge} across all six DOFs with rotational terms weighted by $N_{\alpha\beta}t^{2}/12$, representing a thickness-integrated continuum approximation written directly in displacement gradients.

On pure translations, all four operators match Eq.~\eqref{eq:identity} to a relative error of $8.88 \times 10^{-15}$ because they share the identical membrane block of Eq.~\eqref{eq:kge}.
On pure rotational test vectors, the membrane form yields zero geometric stiffness (relative error $-1.00$), while the continuum truncation exhibits an incorrect sign ($-1.45$ on general rotations and $-1.04$ on bending rotations).
On pure drilling rotations ($\theta_z$ alone), Eq.~\eqref{eq:director2} confirms that the second-order director vanishes whenever bending rotations are zero, leaving $\mathrm{d}^2 W/\mathrm{d}s^2 = 0$.
Nonetheless, the continuum truncation produces a spurious non-zero stiffness ($8.36\times 10^{-2}$), introducing an unphysical rotational stiffening artifact.
Finally, on coupled vectors (the critical buckling mode and a random vector), the block form lacks the translation-rotation couplings of Eq.~\eqref{eq:couplings} and deviates by $-2.18 \times 10^{-3}$ and $-2.33 \times 10^{-4}$, whereas the full form closes to $-2.32 \times 10^{-13}$ and $2.00 \times 10^{-15}$.
Although the missing coupling contribution appears modest ($-2.18 \times 10^{-3}$, or $-0.22\%$) on that single test vector, the shear-weighted coupling shifts structural buckling eigenvalues on design~A by roughly $30\%$ (Section~\ref{sec:res-operator}, Table~\ref{tab:forms}).
Overall, Eq.~\eqref{eq:identity} verifies each analytical operator as an exact representation of its stated truncation; determining which operator commercial solvers adopt requires the comparisons in Section~\ref{sec:results}.

\begin{table}[pos=!htbp]
\centering
\caption{Relative errors $(\vect{\varphi}\T\Kg\vect{\varphi} - W'')/W''$ of assembled stress stiffness operators against Eq.~\eqref{eq:identity} on the $0.25$~m shear panel (absolute values on the drilling-only row where $W'' = 0$).}
\label{tab:identity}
\begin{tabular}{lrrrrr}
\toprule
test vector & $\mathrm{d}^{2}W/\mathrm{d}s^{2}$ & membrane & continuum & block form & full form \\
\midrule
translations only      & $1.702109\times 10^{4}$   & $8.88\times 10^{-15}$  & $8.88\times 10^{-15}$  & $8.88\times 10^{-15}$  & $8.88\times 10^{-15}$ \\
rotations only         & $-1.879945\times 10^{-1}$ & $-1.00$                & $-1.45$                & $-7.61\times 10^{-14}$ & $-7.61\times 10^{-14}$ \\
bending rotations only & $8.953683\times 10^{-1}$  & $-1.00$                & $-1.04$                & $4.00\times 10^{-15}$  & $4.00\times 10^{-15}$ \\
drilling only          & $0$                       & $0$                    & $8.36\times 10^{-2}$   & $0$                    & $0$ \\
\addlinespace
critical mode          & $9.409964\times 10^{1}$   & $-9.95\times 10^{-4}$  & $4.31\times 10^{-3}$   & $-2.18\times 10^{-3}$  & $-2.32\times 10^{-13}$ \\
random, all freedoms   & $4.175978\times 10^{4}$   & $-2.39\times 10^{-4}$  & $-1.98\times 10^{-4}$  & $-2.33\times 10^{-4}$  & $2.00\times 10^{-15}$ \\
\bottomrule
\end{tabular}
\end{table}

\subsection{Cross-code baseline checks and controls}
\label{sec:verif:cross}

Two commercial finite element packages are evaluated alongside our in-house solver: ANSYS Mechanical APDL (SHELL181 with full integration and incompatible modes; eight-node SHELL281) and Abaqus/Standard (S4 full-integration enhanced strain; S4R reduced integration with hourglass control; eight-node S8R).
Every commercial input deck is generated node-for-node from the identical finite element mesh, following the tied-grid and shared-node junction conventions of Section~\ref{sec:panels}.

To confirm that baseline structural stiffnesses match before comparing shear buckling operators, Table~\ref{tab:fulldensity} compares our solver against ANSYS SHELL181 on the $24 \times 8$ dense axial panel.
Variant~2 reproduces full-integration SHELL181 within $0.3\%$ across all six buckling modes and within $0.04\%$ on static compliance $C$.
The compatible baseline element lies $1.31\%$ to $1.95\%$ above full-integration SHELL181, staying well inside the $4.3\%$ spread between SHELL181's own full and reduced integration options.
On a refined ($505\,000$-DOF) mesh of the axial design, our baseline solver reproduces SHELL181 within $0.16\%$ on the leading load factor, with paired skin buckling modes correlating at $\mathrm{MAC} = 0.9999$; on the coarser tied optimization lattice, the compatible-membrane offset is $2.0\%$.

\begin{table}[pos=!htbp]
  \centering
  \caption{Pure axial compression verification on the densely stiffened $24 \times 8$ panel (differences relative to full-integration SHELL181).}
  \label{tab:fulldensity}
  \small
  \begin{tabular}{ccccccc}
    \toprule
    & \multicolumn{2}{c}{present} & \multicolumn{2}{c}{ANSYS SHELL181} & \multicolumn{2}{c}{difference} \\
    \cmidrule(lr){2-3}\cmidrule(lr){4-5}\cmidrule(lr){6-7}
    mode & Baseline & Variant~2 & reduced & full & Base. & Var.~2 \\
    \midrule
    1 & 10.745 & 10.556 & 10.084 & 10.539 & $+1.95\%$ & $+0.16\%$ \\
    2 & 23.617 & 23.349 & 22.844 & 23.312 & $+1.31\%$ & $+0.16\%$ \\
    3 & 95.223 & 93.572 & 89.492 & 93.552 & $+1.79\%$ & $+0.02\%$ \\
    4 & 119.69 & 117.65 & 113.42 & 117.57 & $+1.80\%$ & $+0.07\%$ \\
    5 & 256.27 & 251.89 & 241.65 & 252.41 & $+1.53\%$ & $-0.20\%$ \\
    6 & 301.66 & 296.14 & 286.80 & 297.01 & $+1.57\%$ & $-0.29\%$ \\
    \midrule
    $C$ & 11.633 & 11.700 & 11.862 & 11.704 & $-0.60\%$ & $-0.04\%$ \\
    \bottomrule
  \end{tabular}
\end{table}

On the tied optimization grids under axial compression, where rotational stress stiffness terms vanish, Abaqus S4 ($6.1494$ on the conforming panel, $10.713$ on the axial design) closely matches full-integration SHELL181 ($6.1453$ and $10.7018$).
Similarly, Abaqus S4R ($10.189$) agrees with reduced-integration SHELL181 ($10.217$) within $0.3\%$.
Control checks also verify that secondary modeling options do not account for the large shear discrepancies reported in Section~\ref{sec:results}:
\begin{enumerate}
  \item \textbf{Integration rule:} Under in-plane bending and shear, reduced-integration SHELL181 (\deck{KEYOPT(3)}~$=0$) reproduces the full-integration classic-to-perturbation gaps of Table~\ref{tab:general} closely ($2.39\%$ vs.\ $2.39\%$ on the conforming panel; $14.33\%$ vs.\ $13.98\%$ on design~A). Likewise, Abaqus S4R matches its two procedures and lies only $0.9\%$ and $1.4\%$ below S4.
  \item \textbf{Rib--skin rotational ties:} Releasing all rotational DOFs from rib-to-skin tie equations shifts shear design~A in Abaqus by only $0.16\%$.
  \item \textbf{Solver software version:} All ANSYS analyses executed on Release~19.2 and repeated on Release~2024~R2 agree within $0.0003\%$ (Appendix~\ref{app:repro}).
\end{enumerate}
Collectively, baseline element options, tie idealizations, and solver releases contribute at most $1.4\%$ (typically $<0.3\%$) to cross-code differences, compared with the $14\%$ SHELL181 procedure gap and up to $39\%$ analytical operator spread on shear design~A (Section~\ref{sec:results}).

All commercial comparisons enforce three export controls: every exported element set matches the exact thresholded in-house topology, resultant edge loads act solely on skin nodes (verified automatically pre-export), and compliance is computed as twice the elastic strain energy while matching eigenvalues by magnitude on identical critical branches (Section~\ref{sec:coverage}).

\section{Results}
\label{sec:results}

Three conventions govern all percentage comparisons in this section.
Relative differences of a value $X$ with respect to an explicitly stated reference $Y$ are defined as $100\,(X/Y - 1)\%$.
The gap between the two SHELL181 procedures is defined as $(\lambda_{\mathrm{classic}} - \lambda_{\mathrm{pert}})/\lambda_{\mathrm{classic}}$, and the rotational block contribution is measured relative to the corresponding result without it.
All reported load factors correspond to the critical magnitude $|\lambda|$ on the active spectral branch of Section~\ref{sec:coverage}, with every modal comparison (including all Modal Assurance Criterion, MAC, values) evaluated on that identical branch and at a stated mesh.

An operator sub-block $\mat{B}$ is considered active on the critical buckling mode $\vect{\varphi}$ only if it is both algebraically non-zero and generates non-negligible modal virtual work.
Modal activity is quantified by the quadratic form ratio $|\vect{\varphi}\T\mat{B}\vect{\varphi}|/|\vect{\varphi}\T\Kg\vect{\varphi}|$ and the vector norm ratio $\|\mat{B}\vect{\varphi}\|/\|\Kg\vect{\varphi}\|$, which are both reported wherever matrix presence and modal effect diverge.

\subsection{Operator difference between the two SHELL181 procedures}
\label{sec:verif:tangent}

ANSYS SHELL181 assembles two mathematically distinct stress stiffness operators across its classic linear eigenvalue and linearized perturbation procedures.
The classic procedure includes the moment-weighted rotational block $\mat{K}_{\sigma,\theta\theta}$ of Eq.~\eqref{eq:block}, whereas the perturbation procedure omits this block ($\mat{K}_{\sigma,\theta\theta} \approx \mat{0}$) and acts as a membrane-only operator.
Selective block elimination confirms that omitting $\mat{K}_{\sigma,\theta\theta}$ accounts for essentially the entire $14\%$ to $15\%$ load-factor separation observed on cantilevered panels under in-plane bending and shear.

The classic linear eigenvalue procedure solves Eq.~\eqref{eq:eigproblem} directly about the linear static pre-stress state $\vect{u}$ under the reference load.
In contrast, the linearized perturbation procedure converges a geometrically nonlinear equilibrium state $\vect{u}_\alpha$ under a base pre-load of $\alpha$ times the reference load, re-evaluates element tangent matrices, and solves the perturbed eigenproblem
\begin{equation}
  \left[\mat{K}_{\mathrm{T}}(\vect{u}_\alpha)
  + \beta\,\Kg^{\mathrm{p}}(\vect{u}_\alpha;\Delta\vect{f})\right]\vect{\psi}
  = \vect{0} ,
  \label{eq:twopert}
\end{equation}
where $\mat{K}_{\mathrm{T}}$ is the tangent stiffness and $\Kg^{\mathrm{p}}$ is the geometric stiffness from one application of the perturbation load $\Delta\vect{f}$.
Adding the base pre-load yields the signed total load factor $\alpha + \beta$, with reported critical load-factor magnitude $\lambda_{\mathrm{cr}} = |\alpha + \beta|$ (Section~\ref{sec:coverage}).
Both system pairs were extracted directly from ANSYS solver files via \deck{HBMAT} at $\alpha = 0.01$ with identical DOF ordering.

Table~\ref{tab:tangent} reports the Frobenius norms of the partitioned sub-blocks in the exported stress stiffness matrices across benchmark panels.
In SHELL181, the translational sub-blocks ($\mat{K}_{\sigma,uu}$) agree within $0.1\%$ in Table~\ref{tab:tangent} (and within $0.18\%$ entrywise in Frobenius norm across all exported SHELL181 matrices), whereas the rotation--rotation sub-block ($\mat{K}_{\sigma,\theta\theta}$) has Frobenius norm $10^2$ to $10^3$ in the classic procedure and vanishes to machine precision ($10^{-8}$ to $5\times 10^{-7}$) in perturbation.
Moreover, the exported classic rotational block has zero diagonal entries and zero trace, matching the algebraic structure of Eq.~\eqref{eq:block}.
In contrast, SHELL281 retains the rotational block in both procedures ($6.1920\times 10^2$); its translational norms differ slightly ($3.4\%$ in scalar Frobenius norm, $8.7\%$ entrywise) without shifting the critical load factor.

When compared against our analytical rotational block on the conforming panel, both lack drilling--drilling entries and their off-diagonal drilling-to-bending norms agree within $6\%$, while global rib-to-rib entries differ by a factor of $1.9$ due to finite-rotation parameterization.
Substituting our analytical block into the exported commercial classic generalized eigenproblem reproduces the commercial classic load factor within $0.3\%$ on rebuilt design~A ($9.6685$ vs.\ $9.6413$) and within $0.05\%$ on the conforming panel ($4.8320$ vs.\ $4.8342$).

\begin{table}[pos=tbp]
\centering
\caption{Frobenius norms of partitioned stress stiffness sub-blocks and critical load factors $\lambda$ exported from ANSYS SHELL181 and SHELL281 classic and perturbation procedures.}
\label{tab:tangent}
\begin{tabular}{llrrrr}
\toprule
model & procedure & trans-trans & trans-rot & rot-rot & $\lambda$ \\
\midrule
conforming panel, SHELL181 & classic      & $4.6131\times10^{7}$ & $6.4805\times10^{3}$ & $1.4472\times10^{2}$  & $4.9002$ \\
                           & perturbation & $4.6132\times10^{7}$ & $6.4805\times10^{3}$ & $4.4611\times10^{-7}$ & $4.7670$ \\
\addlinespace
shear A, SHELL181          & classic      & $4.5841\times10^{7}$ & $1.7576\times10^{4}$ & $4.2622\times10^{2}$  & $10.0886$ \\
                           & perturbation & $4.5841\times10^{7}$ & $1.7575\times10^{4}$ & $9.5330\times10^{-8}$ & $8.5901$  \\
\addlinespace
shear A rebuilt, SHELL181  & classic      & $7.5748\times10^{7}$ & $1.7166\times10^{4}$ & $3.0379\times10^{2}$  & $9.6413$ \\
                           & perturbation & $7.5748\times10^{7}$ & $1.7166\times10^{4}$ & $1.2269\times10^{-8}$ & $8.2934$ \\
\addlinespace
shear A rebuilt, SHELL281  & classic      & $1.0703\times10^{8}$ & $3.2628\times10^{4}$ & $6.1920\times10^{2}$  & $6.5937$ \\
                           & perturbation & $1.0339\times10^{8}$ & $3.2628\times10^{4}$ & $6.1920\times10^{2}$  & $6.5940$ \\
\bottomrule
\end{tabular}
\end{table}

Selective block elimination shows that $\mat{K}_{\sigma,\theta\theta}$ accounts for nearly the entire disparity between the two SHELL181 procedures (Table~\ref{tab:removal}).
Zeroing out $\mat{K}_{\sigma,\theta\theta}$ from the exported classic matrix system drops the rebuilt design~A eigenvalue from $9.6413$ to $8.2914$, matching the perturbation load factor ($8.2934$) within $0.03\%$ ($0.02\%$ of the classic value).
Across all tied and rebuilt panels, removing the rotational block accounts for $99.86\%$ to $100.15\%$ of the procedure gap, whereas substituting the perturbation translational block or tangent stiffness alters the load factor by less than $0.03\%$ and $0.24\%$, respectively.

\begin{table}[pos=tbp]
\centering
\caption{Selective elimination of the rotational block $\mat{K}_{\sigma,\theta\theta}$ in exported ANSYS SHELL181 classic matrices compared with perturbation load factors.}
\label{tab:removal}
\begin{tabular}{lrrrrrr}
\toprule
model & rotational rows & classic & block removed & perturbation & share & residual \\
\midrule
conforming panel & $9\,792$ & $4.900236$  & $4.766964$  & $4.767024$  & $100.05\%$ & $-0.0012\%$ \\
shear design A   & $11\,574$ & $10.088644$ & $8.588114$  & $8.590142$  & $100.14\%$ & $-0.0201\%$ \\
shear design B   & $11\,556$ & $10.052638$ & $8.645085$  & $8.643066$  & $99.86\%$  & $+0.0201\%$ \\
combined design  & $11\,538$ & $6.630099$  & $5.782301$  & $5.781386$  & $99.89\%$  & $+0.0138\%$ \\
\addlinespace
\multicolumn{7}{l}{\footnotesize shared-node rebuilds, one subdivision} \\
conforming panel & $24\,192$ & $4.834221$  & $4.718564$  & $4.718537$  & $99.98\%$  & $+0.0006\%$ \\
shear design A   & $27\,756$ & $9.641258$  & $8.291387$  & $8.293350$  & $100.15\%$ & $-0.0204\%$ \\
\bottomrule
\end{tabular}
\\[0.5ex]{\footnotesize share $= (\lambda_{\mathrm{classic}} - \lambda_{\mathrm{removed}})/(\lambda_{\mathrm{classic}} - \lambda_{\mathrm{pert}})$, residual $= (\lambda_{\mathrm{removed}} - \lambda_{\mathrm{pert}})/\lambda_{\mathrm{classic}}$, on the classic critical branch.}
\end{table}

\begin{figure}[tbp]
  \centering
  \includegraphics{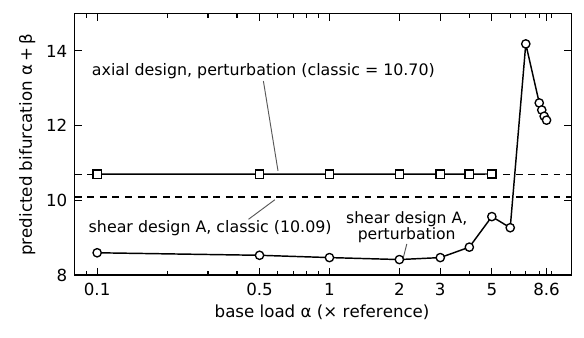}
  \caption{SHELL181 perturbation base-load sweep on tied shear design~A and the axial design: predicted bifurcation load $\alpha + \beta$ (Eq.~\eqref{eq:twopert}) versus forward base-load parameter $\alpha$, alongside classic linear eigenvalue references.}
  \label{fig:ladder}
\end{figure}

Varying the perturbation base-load parameter $\alpha$ across a thirtyfold moderate pre-load range ($\alpha = 0.1$ to $3.0$) confirms that this gap originates from the operator definition rather than nonlinear pre-buckling deformation (Fig.~\ref{fig:ladder}).
Over the moderate pre-load range $\alpha = 0.1$ to $3.0$, the predicted perturbation load factor on design~A remains between $8.41$ and $8.60$ (within $2.2\%$ of its small-load value), well below the tied classic value ($10.09$). At larger pre-loads approaching the nonlinear bifurcation region (shown up to $\alpha = 8.6$ in Fig.~\ref{fig:ladder}), the equilibrium-path response rises sharply; those high-load points are not used to identify the near-zero-load operator. The sweep in Fig.~\ref{fig:ladder} applies the base load in the forward direction; the critical design~A load factor tabulated elsewhere lies on the reversed branch ($0.3\%$ lower at small base load).
Under pure axial compression, where pre-stress bending moments are absent, both procedures yield identical load factors ($10.7018$).

\subsection{Correspondence between commercial elements and analytical operators}
\label{sec:res-operator}

Across cantilevered stiffened panels under in-plane bending and shear, the four commercial shell formulations align closely with distinct analytical stress stiffness operators.
ANSYS SHELL181 classic corresponds to the uncoupled block form ($\mat{K}_{\sigma,\theta\theta}$ active without shear coupling), whereas SHELL181 perturbation omits $\mat{K}_{\sigma,\theta\theta}$ and follows the membrane form.
ANSYS SHELL281 exhibits the same active sub-block structure as the analytical full form (confirmed via matrix export) and Abaqus S4 behaves consistently with that full-form operator group, whereas Abaqus S8R yields markedly lower buckling loads despite sharing the linear elastic stiffness and critical mode of SHELL281.

\subsubsection{Role of individual stress stiffness couplings}

To isolate how each component of the stress stiffness matrix governs buckling capacity, Table~\ref{tab:forms} tracks the critical load factors as coupling terms are added progressively.
Starting from the membrane-only operator ($\lambda = 8.7003$ on tied design~A), introducing the moment-weighted rotational block $\mat{K}_{\sigma,\theta\theta}$ raises the load factor to $10.2577$ ($+17.9\%$ stiffening) due to additional rotational resistance at rib--skin intersections.
Adding the moment--rotation coupling term of Eq.~\eqref{eq:couplings} leaves the eigenvalue virtually unchanged ($\lambda = 10.2526$, $<0.05\%$ shift) because its modal quadratic form on the critical mode vanishes to five decimal places.
In contrast, adding the transverse shear coupling term $\mat{K}_{\sigma,u\theta}$ drops the buckling load factor from $10.2526$ to $7.1328$ (a $30.4\%$ reduction).
Under in-plane bending and shear, in-plane rib bending generates substantial transverse shear forces along the stiffener walls that couple with out-of-plane rotations across non-coplanar junctions, supplying destabilizing virtual work that dominates the four-node element separation.

\begin{table}[pos=tbp]
\centering
\caption{Critical buckling load factors of the analytical stress stiffness formulations on the tied designs, evaluating the individual effect of moment- and shear-weighted couplings.}
\label{tab:forms}
\begin{tabular}{llrrr}
\toprule
Formulation & Included matrix terms & shear A & shear B & combined \\
\midrule
membrane                & Eq.~\eqref{eq:kge}                                                               & $8.7003$  & $8.7565$  & $5.8700$ \\
block                   & Eq.~\eqref{eq:kge} + Eq.~\eqref{eq:block}                                        & $10.2577$ & $10.2194$ & $6.7411$ \\
block + moment coupling & Eq.~\eqref{eq:kge} + Eq.~\eqref{eq:block} + $M$ term of Eq.~\eqref{eq:couplings} & $10.2526$ & $10.2138$ & $6.7402$ \\
full                    & Eq.~\eqref{eq:kge} + Eq.~\eqref{eq:block} + Eq.~\eqref{eq:couplings}             & $7.1328$  & $7.2457$  & $5.0684$ \\
\bottomrule
\end{tabular}
\end{table}

\subsubsection{Classification of ANSYS and Abaqus shell elements}

Tables~\ref{tab:map} and~\ref{tab:levels} compare the analytical operators against commercial finite element predictions across tied models and shared-node rebuilds, respectively; among the other three tabulated six-DOF commercial shell elements (S4, SHELL281, and S8R), classic and perturbation procedures agree within $0.06\%$.
For ANSYS SHELL181 and SHELL281, the underlying operators are identified directly by inspecting the exported global stiffness and stress stiffness matrices.
In SHELL181, the classic procedure tracks our analytical block form within $0.16\%$ to $1.65\%$ on the tied shear and combined panels and within $0.50\%$ to $1.50\%$ across all shared-node rebuilds (Tables~\ref{tab:map}--\ref{tab:levels}), while the perturbation procedure tracks the membrane form within $0.12\%$ to $1.51\%$.
Under pure axial compression, where bending moments vanish, all three analytical forms and both SHELL181 procedures coincide ($10.7321$ vs.\ $10.5714$ on the level-1 rebuild).
The $1.5\%$ baseline offset under axial loading reflects compatible-membrane kinematics. On tied shear design~A (Table~\ref{tab:map}), adopting Variant~2 (Section~\ref{sec:femodel}) reduces the remaining SHELL181 discrepancies to $0.57\%$ for classic versus block and $0.12\%$ for perturbation versus membrane.
Crucially, although SHELL181 exports a non-zero translation--rotation block, its quadratic form on the critical mode is negligible ($+0.03\%$ on design~A), confirming that SHELL181 exhibits no active shear--rotation coupling.

In contrast, direct matrix evidence classifies ANSYS SHELL281 within the full-form operator group, with the same active sub-block structure as the analytical full form.
Across its classic and perturbation procedures, the exported rotational sub-block $\mat{K}_{\sigma,\theta\theta}$ agrees within $0.12\%$ (and the translation--rotation sub-block within $0.14\%$) in Frobenius norm, yielding load factors that match within $0.005\%$ despite an $8.7\%$ entrywise difference in the translational sub-block.
Selective matrix zeroing on rebuilt design~A directly exposes its sub-blocks: zeroing $\mat{K}_{\sigma,\theta\theta}$ lowers the load factor by $14.2\%$ (reproducing the $14.0\%$ gap between SHELL181's classic and perturbation procedures), whereas zeroing $\mat{K}_{\sigma,u\theta}$ raises the load factor by $34.5\%$ (recovering the stiffening behavior of the block form). Zeroing both blocks simultaneously yields $\lambda = 7.7315$, close to the converged level-3 four-node values of SHELL181 perturbation ($7.7992$) and the analytical membrane form ($7.8863$).
As a result, converged SHELL281 predictions ($6.5983$ on design~A) lie within $2.3\%$ of our analytical full form ($6.7531$).

\begin{table}[pos=tbp]
\centering
\caption{Critical buckling load factors of analytical operators and commercial four-node shell formulations on the tied panel models (shared-node rebuilds and 3D continuum benchmarks follow in Table~\ref{tab:levels}).}
\label{tab:map}
\small
\begin{tabular}{lrrrrrr}
\toprule
 & \multicolumn{3}{c}{Analytical forms} & \multicolumn{2}{c}{SHELL181} & S4 \\
\cmidrule(lr){2-4}\cmidrule(lr){5-6}
Model & Membrane & Block & Full & Classic & Pert. & \\
\midrule
conforming panel$^{\dagger}$ & $4.7765$ & $4.9081$ & $4.6494$ & $4.9002$ & $4.7670$ & $4.6365$ \\
shear A       & $8.7003$ & $10.2577$ & $7.1328$ & $10.0886$ & $8.5901$ & $7.0236$ \\
shear A$^{\dagger}$ & $8.6002$ & $10.1464$ & $7.0363$ & $10.0886$ & $8.5901$ & $7.0236$ \\
shear B       & $8.7565$ & $10.2194$ & $7.2457$ & $10.0526$ & $8.6431$ & $7.1357$ \\
combined      & $5.8700$ & $6.7411$  & $5.0684$ & $6.6301$  & $5.7814$ & $4.9882$ \\
axial         & $10.9121$ & $10.9121$ & $10.9121$ & $10.7018$ & $10.7018$ & $10.713$ \\
\bottomrule
\end{tabular}
\\[0.5ex]{\footnotesize $^{\dagger}$ Evaluated under Variant~2 of Section~\ref{sec:femodel} (incompatible modes and Hughes--Brezzi penalty). On the symmetric conforming panel, forward and reversed SHELL181 perturbation branches agree within $0.002\%$ ($4.7670$ vs.\ $4.7669$).}
\end{table}

\begin{figure}[tbp]
  \centering
  \includegraphics[width=\linewidth]{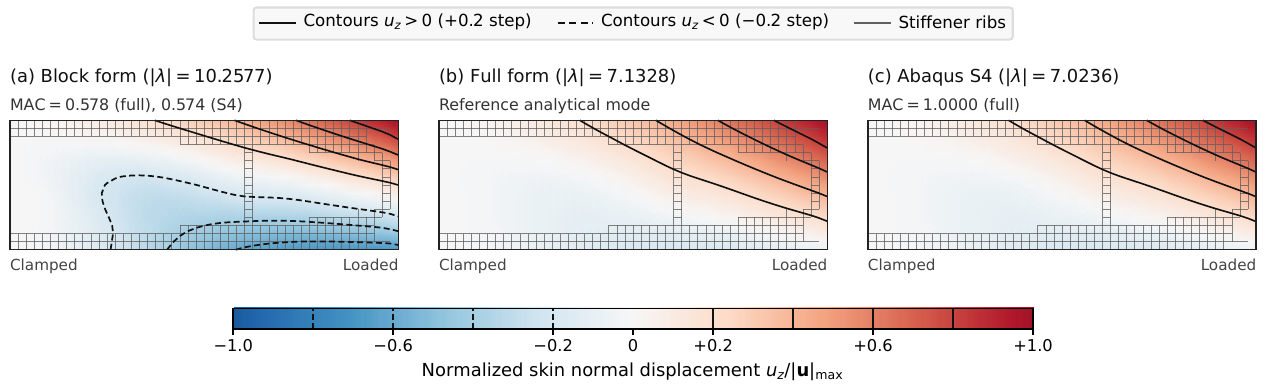}
  \caption{Critical buckling modes of tied shear design~A (reversed branch) in plan view: (a)~block form, (b)~full form, and (c)~Abaqus S4, shaded by normalized skin normal displacement $u_z/\|\vect{u}\|_{\max}$ with stiffener ribs in gray.}
  \label{fig:modes}
\end{figure}

Because Abaqus does not export internal shell stress stiffness matrices, Abaqus S4 is classified through load-factor tracking, elastic strain energy checks, and modal assurance.
The Abaqus S4 linear eigenvalue and perturbation procedures agree within $0.03\%$ on the tied and level-1--2 panels and within $0.06\%$ at level~3. Across the benchmarks, S4 tracks the analytical full form within $0.18\%$ to $1.58\%$ on the tied shear and combined panels and within $1.0\%$ to $1.7\%$ on the rebuilt shear panels (Tables~\ref{tab:map}--\ref{tab:levels}).
Benchmarked against the exported SHELL181 stiffness on the same nodal grid, the small-displacement elastic strain energy on the critical mode agrees within $0.3\%$ on translations ($1.8\%$ on design~A and $7.3\%$ on the conforming panel including bending rotations).
The translational MAC between S4 and the analytical full form is $1.0000$ across all tested panels. By contrast, the block-form MAC drops to $0.578$ against the full form and $0.574$ against S4. The block form therefore produces an artificial torsional twist, whereas S4 and the full form produce the same localized corner-lift skin mode (Fig.~\ref{fig:modes}).

In contrast to SHELL281 and S4, Abaqus S8R yields markedly lower buckling loads on stiffened shear panels, converging to $\lambda = 5.5955$ on rebuilt design~A ($15.2\%$ below SHELL281; Table~\ref{tab:levels}).
This difference does not stem from linear elastic stiffness: S8R shares the critical mode of SHELL281 ($\mathrm{MAC} = 0.991$) and agrees in linear elastic strain energy on that mode within $0.3\%$, whereas its effective stress stiffness supplies $20\%$ more destabilizing work on design~A.
Testing the five-DOF element Abaqus S8R5 gives a similarly low classic load factor ($\lambda = 5.6507$) together with a $5.3\%$ classic-to-perturbation drop. Drilling treatments at fold lines are therefore unlikely to be the primary cause; the discrepancy is more plausibly associated with S8R's internal stress stiffness assembly.

\subsubsection{Mesh convergence across shell formulations}

\begin{figure*}[tbp]
  \centering
  \includegraphics{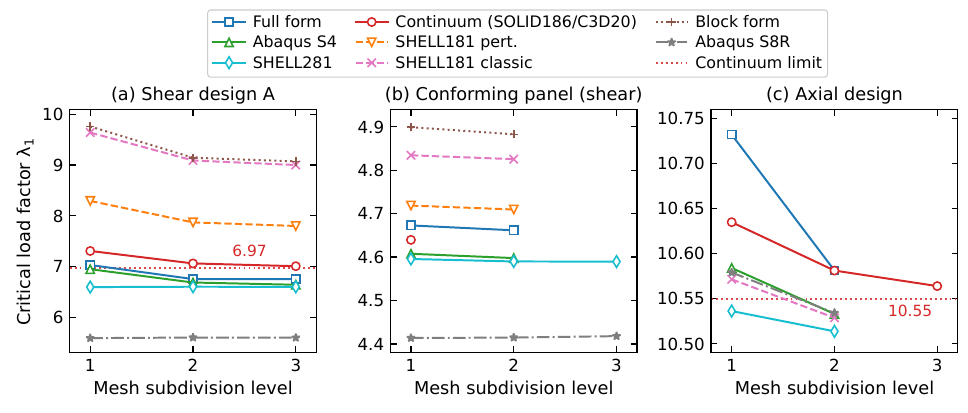}
  \caption{Mesh convergence of critical buckling load factors across subdivision levels for: (a) shear design~A; (b) conforming panel under shear; (c) axial design. The dotted red line indicates the estimated Richardson-extrapolated continuum limit.}
  \label{fig:meshconv}
\end{figure*}

Table~\ref{tab:levels} and Fig.~\ref{fig:meshconv} track mesh refinement across three subdivision levels for rebuilt design~A, the conforming panel, and the axial benchmark. Subdivision levels~1, 2, and~3 place $1$, $4$, and $9$ eight-node elements (or $4$, $16$, and $36$ four-node elements) per design cell on identical node lattices, while the 3D continuum reference runs on its own coarse, medium, and fine meshes.
On shear design~A, refinement from level~1 to level~3 stabilizes predictions across every operator: the full form converges from $7.0292$ to $6.7531$ (changing by $<0.02\%$ from level~2 to~3), Abaqus S4 reaches $6.6397$, and ANSYS SHELL281 reaches $6.5983$.
Crucially, mesh refinement preserves the relative separation among operators: the SHELL181 procedure gap stays at $14.0\%$ at level~1 and $13.4\%$ at level~3, while S4 tracks the full form within $1.0\%$ to $1.7\%$ throughout.
Under pure axial compression, all four commercial elements and all three analytical forms converge to $10.51$ to $10.58$ (within $0.7\%$), supporting the conclusion that the shear disparities arise primarily from the stress stiffness operators rather than from mesh discretization.

\begin{table}[pos=tbp]
\centering
\caption{Mesh convergence of critical buckling load factors across subdivision levels on shared-node rebuilds and the three 3D continuum meshes (dashes denote unevaluated levels).}
\label{tab:levels}
\small
\begin{tabular}{lrrrrrrrrr}
\toprule
 & \multicolumn{3}{c}{shear design A} & \multicolumn{3}{c}{conforming panel} & \multicolumn{3}{c}{axial design} \\
\cmidrule(lr){2-4}\cmidrule(lr){5-7}\cmidrule(lr){8-10}
Formulation & $1$ & $2$ & $3$ & $1$ & $2$ & $3$ & $1$ & $2$ & $3$ \\
\midrule
membrane form     & $8.3763$ & $7.9223$ & $7.8863$ & $4.7840$ & $4.7703$ & -- & $10.7321$ & $10.5812$ & -- \\
block form        & $9.7628$ & $9.1449$ & $9.0729$ & $4.8989$ & $4.8825$ & -- & $10.7321$ & $10.5812$ & -- \\
full form         & $7.0292$ & $6.7518$ & $6.7531$ & $4.6728$ & $4.6616$ & -- & $10.7320$ & $10.5812$ & -- \\
SHELL181 classic  & $9.6413$ & $9.0906$ & $9.0011$ & $4.8342$ & $4.8253$ & -- & $10.5714$ & $10.5287$ & -- \\
SHELL181 pert.    & $8.2934$ & $7.8699$ & $7.7992$ & $4.7185$ & $4.7095$ & -- & $10.5715$ & $10.5287$ & -- \\
S4                & $6.9467$ & $6.6839$ & $6.6397$ & $4.6076$ & $4.5977$ & -- & $10.584$ & $10.533$ & -- \\
SHELL281          & $6.5937$ & $6.6009$ & $6.5983$ & $4.5956$ & $4.5899$ & $4.5896$ & $10.5362$ & $10.5137$ & -- \\
S8R               & $5.5872$ & $5.5987$ & $5.5955$ & $4.4138$ & $4.4148$ & $4.4181$ & $10.579$ & $10.534$ & -- \\
continuum         & $7.3062$ & $7.0606$ & $7.0058$ & $4.6396$ & -- & -- & $10.6348$ & $10.5811$ & $10.5639$ \\
\bottomrule
\end{tabular}
\end{table}

\subsection{Continuum benchmark and mesh sensitivity}
\label{sec:res-continuum}

Independent 3D 20-node hexahedral continuum models (ANSYS SOLID186 and Abaqus C3D20) converge monotonically to $\lambda = 7.0058$ on rebuilt shear design~A (Richardson-extrapolated limit $\approx 6.97$), with both solvers agreeing within $0.002\%$.
Through-thickness refinement from two to four hexahedral layers shifts the load factor by only $-0.41\%$ to $-0.45\%$, indicating that the remaining through-thickness discretization effect is small relative to the shell-operator differences.
For the baseline junction geometry, the \{full form, Abaqus S4, ANSYS SHELL281\} group lies closest to the continuum reference ($3.6\%$ to $5.8\%$ below continuum), though junction-idealization sensitivity ($\approx 2.5\%/\mathrm{mm}$ under shear; $2.7\%/\mathrm{mm}$ under axial compression) prevents distinguishing among its three members.

\subsubsection{In-plane and through-thickness mesh convergence}

To obtain a physical reference free of shell director kinematics and drilling regularizations, 3D continuum models were constructed with 20-node quadratic hexahedral elements (SOLID186 and C3D20).
In the baseline geometry, the $10$~mm skin is centered on the shell mid-surface, while $8$~mm ribs stand directly on the skin's top face to reach a total height of $0.100$~m with shared interface nodes.
On the axial compression design, where out-of-plane junction bending is absent, the continuum converges monotonically across three meshes to $\lambda = 10.5639$ (estimated Richardson limit $10.55$), matching all four commercial shell formulations ($10.5137$ to $10.534$, Table~\ref{tab:levels}) within $0.3\%$ to $0.5\%$.

On rebuilt shear design~A, in-plane continuum refinement reduces the buckling load factor from $7.3062$ (coarse) to $7.0606$ (medium) and $7.0058$ (fine). ANSYS and Abaqus agree within $0.002\%$ on the fine mesh (Table~\ref{tab:ribfoot}). The medium-to-fine change is $0.8\%$, and Richardson extrapolation gives a limiting value of $6.97$.
To verify that through-thickness discretization does not bias this sequence, skin and rib thicknesses were refined independently on both the coarse and medium in-plane grids (Table~\ref{tab:thickness}).
On the coarse layout, increasing through-thickness layers from $1$ to $2$ and $4$ quadratic elements ($3$, $5$, and $9$ nodes through depth) scales the model from $13\,765$ to $161\,824$ elements ($92\,168$ to $759\,713$ nodes) and converges monotonically from $7.3062$ to $7.2043$ ($-1.39\%$) and $7.1749$ ($-0.41\%$).
Refining the medium layout similarly from $2$ to $4$ layers ($240\,680$ elements and $1\,127\,951$ nodes) gives $\lambda = 7.0286$ ($-0.45\%$), within $0.33\%$ of the fine 2-layer baseline ($7.0058$) and $0.84\%$ of the extrapolated limit ($6.97$).
Four-layer refinement on the fine mesh ($4.71\times 10^6$ DOFs, estimated $72.9$~GB in-core RAM) was attempted but terminated because in-core memory and scratch requirements exceeded available 64~GB workstation resources, leaving the four-layer medium grid as the finest completed four-layer model. Nevertheless, the consistent $\approx -0.4\%$ shifts on both coarse and medium grids indicate that the remaining through-thickness discretization effect is small relative to the shell-operator differences.

\begin{table}[pos=tbp]
\centering
\caption{Through-thickness refinement of the 3D continuum reference (ANSYS SOLID186) on coarse and medium in-plane meshes of shear design~A.}
\label{tab:thickness}
\footnotesize
\setlength{\tabcolsep}{6pt}
\begin{tabular}{lcrrc}
\toprule
In-plane mesh & \makecell{Layers in\\thickness} & \makecell{SOLID186\\elements} & \makecell{Total\\nodes} & \makecell{Buckling load\\factor $\lambda$} \\
\midrule
Coarse layout & 1 & $13\,765$ & $92\,168$ & $7.3062$ \phantom{($-0.00\%$)} \\
              & 2 & $41\,992$ & $224\,785$ & $7.2043$ ($-1.39\%$) \\
              & 4 & $161\,824$ & $759\,713$ & $7.1749$ ($-0.41\%$) \\
\addlinespace
Medium layout & 2 & $71\,956$ & $385\,971$ & $7.0606$ \phantom{($-0.00\%$)} \\
              & 4 & $240\,680$ & $1\,127\,951$ & $7.0286$ ($-0.45\%$) \\
\bottomrule
\end{tabular}
\end{table}

\subsubsection{Rib--skin junction sensitivity and formulation grouping}

At rib--skin intersections, the half-skin thickness offset (Section~\ref{sec:panels}, Fig.~\ref{fig:panel}d) constitutes the main geometric idealization difference between shell mid-surface models ($100$~mm blade height) and the baseline solid model ($95$~mm blade above the top skin face).
Raising the rib top by $5$~mm to expose a full $100$~mm blade above the solid skin raises bending stiffness and increases continuum buckling loads by $12.6\%$ to $13.4\%$ across axial and shear cases, with ANSYS and Abaqus agreeing within $0.014\%$ (Table~\ref{tab:ribfoot}).

\begin{figure}[tbp]
  \centering
  \includegraphics{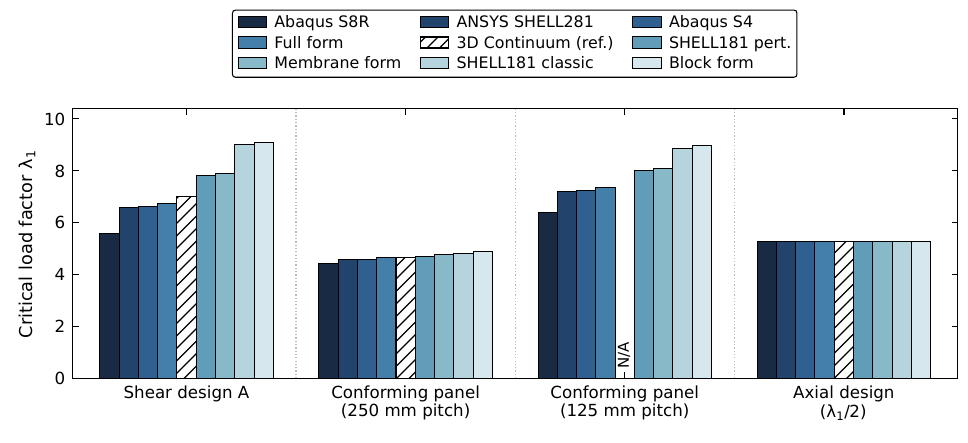}
  \caption{Comparison of critical buckling load factors on the finest shared-node meshes across analytical operators, commercial shell elements, and the 3D continuum reference (hatched); axial-design values are halved ($\lambda_1/2$) for scale.}
  \label{fig:summary}
\end{figure}

Under pure axial compression, the baseline solid matches fine shells within $0.3\%$ to $0.5\%$ whereas the raised-rib model overpredicts by $14.5\%$ to $14.7\%$, motivating the baseline junction geometry.
Because the rotational block contributes negligibly under axial compression, however, axial calibration does not remove junction-idealization uncertainty under shear ($\approx 2.5\%/\mathrm{mm}$, versus $2.7\%/\mathrm{mm}$ under axial compression), so Table~\ref{tab:ribfoot} retains the raised-rib configuration as an explicit bracket.
Against the baseline fine continuum ($\lambda = 7.0058$, extrapolated limit $6.97$), the analytical full form ($3.6\%$ below), Abaqus S4 ($5.2\%$ below), and ANSYS SHELL281 ($5.8\%$ below) lie closest, whereas SHELL181 perturbation ($11.3\%$ above) and SHELL181 classic ($28.5\%$ above) overpredict and Abaqus S8R underpredicts substantially ($20.1\%$ below; Fig.~\ref{fig:summary}).
Because predictions within the \{full form, S4, SHELL281\} group span only $2.3\%$, the continuum benchmark cannot distinguish among them and none is claimed uniquely correct.

\begin{table}[pos=tbp]
\centering
\caption{Sensitivity of 3D continuum buckling load factors (SOLID186, C3D20) to rib--skin junction height, compared with finest shell predictions from Table~\ref{tab:levels} (dashes denote unevaluated meshes).}
\label{tab:ribfoot}
\small
\begin{tabular}{llrrrr}
\toprule
 & & \multicolumn{2}{c}{baseline (as built)} & \multicolumn{2}{c}{rib raised $5$~mm} \\
\cmidrule(lr){3-4}\cmidrule(lr){5-6}
Model & Mesh & SOLID186 & C3D20 & SOLID186 & C3D20 \\
\midrule
axial design & coarse  & $10.6348$ & $10.632$ & $12.0572$ & $12.056$ \\
             & medium  & $10.5811$ & $10.581$ & -- & -- \\
             & fine    & $10.5639$ & -- & -- & -- \\
             & shells  & \multicolumn{4}{l}{$10.5137$ to $10.534$ (SHELL281, SHELL181, S4, S8R)} \\
\addlinespace
shear design A & coarse & $7.3062$ & $7.3040$ & $8.2351$ & $8.2340$ \\
             & medium  & $7.0606$ & $7.0603$ & $7.9522$ & $7.9520$ \\
             & fine    & $7.0058$ & $7.0059$ & -- & -- \\
             & shells  & \multicolumn{4}{l}{$6.7531$ full form, $6.6397$ S4, $6.5983$ SHELL281} \\
\bottomrule
\end{tabular}
\end{table}

\subsubsection{Nonlinear pre-buckling path, rotation parameterization, and layout sensitivity}

Tracing geometrically nonlinear pre-buckling equilibrium paths up to eleven times the reference reversed shear load on design~A shows that operator separation is not an artifact of linearizing near zero load.
Across moderate pre-loads ($\alpha = 1$ to $3$), linearized perturbation predictions ($\alpha + \beta$) stay cleanly separated ($8.15$ to $8.16$ for SHELL181, $6.5$ to $6.8$ for S4, $6.1$ to $6.4$ for SHELL281, and $4.9$ to $5.4$ for S8R; at $\alpha = 5$, SHELL181 rises to $9.13$ while S4 and SHELL281 remain at $6.42$ and $6.26$).
Varying the rotation-parameterization factor $\alpha_{\mathrm{rot}}$ from Section~\ref{sec:buckling} confirms the expected non-uniqueness: relative to the fine continuum ($7.0058$), $\alpha_{\mathrm{rot}} = 0.5$ (exponential map) yields $\lambda = 6.7531$ ($3.6\%$ below), $\alpha_{\mathrm{rot}} = 0$ yields $5.7851$ ($17.4\%$ below), and $\alpha_{\mathrm{rot}} = 1.0$ yields $7.8793$ ($12.5\%$ above).
Finally, comparing shear design~A (block-optimized) with design~B (membrane-optimized) shows that although operators shift buckling loads by $17.9\%$ to $30.4\%$, each layout outperforms the other by only $0.37\%$ and $0.65\%$ under its own operator (Fig.~\ref{fig:designs}). Topological rib layouts therefore remain robust despite large operator sensitivity in predicted load factors.

\subsection{Structural conditions that activate operator discrepancies}
\label{sec:res-when}

Across twelve thin-walled benchmark configurations, substantial formulation divergence activates only on cantilevered stiffened panels subjected to in-plane bending and shear.
On unbranched shells (cylinders and open beams) and pure membrane load states (axial compression and pure shear flow), all three analytical forms coincide and operator discrepancies remain negligible.
Three physical conditions---non-coplanar junctions, junction pre-stress bending or shear, and modal junction drilling---govern this activation.

\subsubsection{Cross-configuration comparison}

Table~\ref{tab:general} compiles critical load factors across all twelve benchmark structures on level-1 shared-node rebuilds, dividing cleanly into three groups:
\begin{enumerate}
  \item \textbf{Dense topologically optimized shear panels:} On shear design~A, the analytical block and membrane forms exceed the full form by $38.9\%$ and $19.2\%$, respectively, while the two SHELL181 procedures diverge by $14.0\%$.
  \item \textbf{Conventionally stiffened panels under in-plane bending and shear:} On the conventional $250$~mm-pitch panel, the analytical forms span $4.8\%$ and SHELL181 procedures diverge by $2.4\%$. Cutting the ribs (36 free rib ends) leaves both gaps nearly unchanged ($4.5\%$ and $2.3\%$), whereas halving the pitch to $125$~mm widens them to $22.2\%$ and $9.7\%$.
  \item \textbf{Unbranched shells and pure membrane loading:} Across the remaining eight cases---pure shear flow, axial compression, four cylindrical cases, and two open beams---analytical forms agree within $0.05\%$ ($0.12\%$ for shear flow), and procedure gaps stay within $0.06\%$ across all non-SHELL181 commercial elements and within $0.04\%$ in SHELL181 (except $-0.27\%$ on the channel cantilever). Remaining cross-element differences (typically $<0.8\%$, reaching $1.1\%$ to $4.6\%$ on coarser cylindrical meshes, $2.8\%$ on the channel cantilever, and $7.1\%$ on the pure shear-flow panel) reflect standard element interpolation rather than stress stiffness truncations.
\end{enumerate}

\begin{table}[pos=tbp]
\centering
\caption{Critical buckling load factors across twelve thin-walled benchmarks on level-1 shared-node rebuilds (classic commercial procedures; SHELL181 gap is classic versus perturbation).}
\label{tab:general}
\small
\begin{tabular}{lrrrrrrrr}
\toprule
 & \multicolumn{3}{c}{Analytical forms} & \multicolumn{2}{c}{SHELL181} & S4 & SHELL281 & S8R \\
\cmidrule(lr){2-4}\cmidrule(lr){5-6}
Configuration & Membrane & Block & Full & Classic & Gap & & & \\
\midrule
conforming panel, shear & $4.7840$ & $4.8989$ & $4.6728$ & $4.8342$ & $2.39\%$ & $4.6076$ & $4.5956$ & $4.4138^{\dagger}$ \\
the same, ribs cut      & $3.2278$ & $3.2994$ & $3.1584$ & $3.2563$ & $2.25\%$ & $3.1158$ & $3.0957$ & $2.9979$ \\
the same, $125$~mm pitch & $8.1049$ & $8.9681$ & $7.3410$ & $8.8555$ & $9.69\%$ & $7.2311$ & $7.1964$ & $6.3989$ \\
the same, shear flow    & $346.57$ & $346.21$ & $346.17$ & $345.77$ & $0.02\%$ & $345.49$ & $321.30$ & $323.79$ \\
shear design A          & $8.3763$ & $9.7628$ & $7.0292$ & $9.6413$ & $13.98\%$ & $6.9467$ & $6.5937$ & $5.5872^{\dagger}$ \\
axial design            & $10.7321$ & $10.7321$ & $10.7320$ & $10.5714$ & $0.00\%$ & $10.584$ & $10.5362$ & $10.579$ \\
\addlinespace
cylinder, axial         & $1.5150$ & $1.5150$ & $1.5150$ & $1.5203$ & $0.00\%$ & $1.5188$ & $1.4846$ & $1.4836$ \\
cylinder, torsion       & $1.6870$ & $1.6870$ & $1.6870$ & $1.6905$ & $0.02\%$ & $1.6903$ & $1.6728$ & $1.6735$ \\
stiffened, axial        & $2.8583$ & $2.8579$ & $2.8577$ & $2.8747$ & $0.02\%$ & $2.8638$ & $2.7413$ & $2.7514$ \\
stiffened, torsion      & $5.2480$ & $5.2482$ & $5.2478$ & $5.2577$ & $0.04\%$ & $5.2540$ & $5.0810$ & $5.0950$ \\
\addlinespace
I beam, end moments     & $4.6049$ & $4.6049$ & $4.6038$ & $4.5999$ & $0.00\%$ & $4.5969$ & $4.5933$ & $4.5815$ \\
channel, tip load       & $7.1848$ & $7.1872$ & $7.1869$ & $7.1230$ & $-0.27\%$ & $7.1248$ & $7.0388$ & $6.9240$ \\
\bottomrule
\end{tabular}
\\[0.5ex]{\footnotesize Evaluated with one eight-node or four four-node elements per design cell; $^{\dagger}$ At three subdivisions S8R reaches $4.4181$ and $5.5955$, while SHELL281 reaches $4.5896$ and $6.5983$.}
\end{table}

\subsubsection{Pre-stress, modal drilling, and junction-density diagnostics}

Because $\mat{K}_{\sigma,\theta\theta}$ and $\mat{K}_{\sigma,u\theta}$ are weighted by bending moments and transverse shear resultants, a pre-stress state lacking these resultants cannot activate either term.
Table~\ref{tab:junction} summarizes element median norms of membrane forces ($\bar{N}$), bending moments ($\bar{M}$), and transverse shear resultants ($Q$) across the twelve shared-node benchmarks alongside modal drilling engagement measured by
\begin{equation}
  r_{\mathrm{d}} = \left(\sum_{e}\sum_{a} \theta_{z,ea}^{2} \Big/
                        \sum_{e}\sum_{a} (\theta_{x,ea}^{2} + \theta_{y,ea}^{2})\right)^{1/2},
  \label{eq:rd}
\end{equation}
evaluated over local DOFs at every node $a$ of element $e$. In Table~\ref{tab:junction}, junction elements contain at least two nodes where adjacent surface normals diverge by more than $30^\circ$, the modal diagnostics ($r_{\mathrm{d}}$, $q$, $n$; Section~\ref{sec:results}) use the membrane-operator critical mode, and the block shift gives the percentage load-factor change from adding $\mat{K}_{\sigma,\theta\theta}$. Supplementary diagnostics for the tied optimization panels and the full-scale box girder are reported in Appendix~\ref{app:repro} (Table~\ref{tab:when}).

Across the benchmarks, stress stiffness discrepancies activate whenever three physical conditions hold simultaneously:
\begin{enumerate}
  \item \textbf{Kinematic intersection:} Non-coplanar structural junctions must exist where the drilling rotation of a stiffener acts as an out-of-plane bending rotation of the skin; in unbranched shells, drilling decouples from bending curvatures.
  \item \textbf{Pre-stress bending and shear:} Substantial bending moments ($6\bar{M}/(\bar{N}t) > 0.5$) or transverse shear forces ($Q/\bar{N} > 0.02$) must develop along the junction footprint. Under axial compression, junction bending remains at $2.3\%$--$2.9\%$ of membrane stress on the rebuild ($1.5\%$ tied), keeping the rotational block inert despite strong modal drilling ($r_{\mathrm{d}} = 1.202$).
  \item \textbf{Modal participation at junctions:} The critical mode must involve substantial junction drilling rotations ($r_{\mathrm{d}} > 0.5$, modal quadratic fraction $q = |\vect{\varphi}\T\mat{B}\vect{\varphi}|/|\vect{\varphi}\T\Kg\vect{\varphi}| > 2\%$). The pure shear-flow panel illustrates this cleanly: despite high junction bending ($6\bar{M}/(\bar{N}t) = 23.779$), localized skin shear lobing leaves junction drilling weak ($r_{\mathrm{d}} = 0.293, q = 0.01\%$) and the rotational block shift at $-0.10\%$.
\end{enumerate}

\begin{table}[pos=tbp]
\centering
\caption{Pre-stress resultant ratios, eigenmode drilling diagnostics, and rotational-block shifts across the twelve shared-node benchmarks.}
\label{tab:junction}
\footnotesize
\setlength{\tabcolsep}{2.0pt}
\begin{tabular}{lrrrrrrrrrr}
\toprule
 & \multicolumn{2}{c}{Elements} & \multicolumn{2}{c}{All elements} & \multicolumn{2}{c}{Junction elements} & \multicolumn{3}{c}{Mode diagnostics} & Block \\
\cmidrule(lr){2-3}\cmidrule(lr){4-5}\cmidrule(lr){6-7}\cmidrule(lr){8-10}\cmidrule(lr){11-11}
Configuration & Total & Junction & $6\bar M/(\bar N t)$ & $Q/\bar N$ & $6\bar M/(\bar N t)$ & $Q/\bar N$ & $r_{\mathrm{d}}$ & $q$ (\%) & $n$ (\%) & shift \\
\midrule
shear design A      & $10\,764$ & $8\,965$ & $0.771$ & $0.0492$ & $1.239$ & $0.0877$ & $1.104$ & $17.0$ & $1.8$ & $+16.55\%$ \\
conforming panel    & $8\,448$  & $3\,360$ & $0.899$ & $0.0158$ & $0.808$ & $0.0231$ & $0.914$ & $2.4$ & $0.6$ & $+2.40\%$ \\
the same, ribs cut  & $8\,232$  & $3\,211$ & $0.715$ & $0.0134$ & $0.685$ & $0.0207$ & $0.888$ & $2.2$ & $0.5$ & $+2.22\%$ \\
the same, $125$~mm pitch & $13\,056$ & $8\,128$ & $1.163$ & $0.0652$ & $1.017$ & $0.0562$ & $1.117$ & $10.0$ & $1.2$ & $+10.65\%$ \\
conforming panel, shear flow & $8\,448$ & $3\,360$ & $40.874$ & $1.3816$ & $23.779$ & $1.3480$ & $0.293$ & $0.01$ & $0.005$ & $-0.10\%$ \\
channel, tip load   & $6\,400$  & $640$     & $0.803$ & $0.0044$ & $0.334$ & $0.0437$ & $0.107$ & $0.03$ & $0.12$ & $+0.03\%$ \\
axial design        & $10\,752$ & $6\,616$ & $0.023$ & $0.0005$ & $0.029$ & $0.0014$ & $1.202$ & $0.00$ & $0.02$ & $-0.00\%$ \\
stiffened cylinder, torsion & $20\,736$ & $6\,912$ & $0.018$ & $0.0027$ & $0.037$ & $0.0029$ & $0.513$ & $0.00$ & $0.002$ & $+0.00\%$ \\
stiffened cylinder, axial   & $20\,736$ & $6\,912$ & $0.024$ & $0.0012$ & $0.036$ & $0.0030$ & $0.614$ & $0.01$ & $0.005$ & $-0.02\%$ \\
I beam, moment      & $20\,480$ & $1\,920$ & $0.000$ & $0.0000$ & $0.030$ & $0.0000$ & $0.159$ & $0.00$ & $0.002$ & $+0.00\%$ \\
plain cylinder, torsion & $12\,288$ & $0$ & $0.007$ & $0.0000$ & -- & -- & $0.222$ & $0.00$ & $0.000$ & $+0.00\%$ \\
plain cylinder, axial   & $12\,288$ & $0$ & $0.000$ & $0.0000$ & -- & -- & $0.402$ & $0.00$ & $0.000$ & $-0.00\%$ \\
\bottomrule
\end{tabular}
\end{table}

Inside any smooth planar plate in static equilibrium without surface couples, the virtual work of constant pre-stress moments over second-order in-plane director increments vanishes identically. Testing the rotational block on an equilibrated constant-moment field ($u=v=0$, $w=-\frac{1}{2}(ax^2+by^2)$, $\theta_x=-by$, $\theta_y=ax$) yields $-8\times 10^{-12}$ across four meshes.
In the present benchmarks, the modal work of both the rotational block and the transverse shear coupling is concentrated at boundaries and non-coplanar junctions. On shear design~A, the skin interior contributes virtually zero virtual work, whereas stiffener walls supply the entire sum ($21\%$ from the bottom rib layer and $79\%$ from upper layers for the $+0.38$ shear-coupling quadratic form).
As the junction element fraction grows from $40\%$ ($250$~mm pitch) to $62\%$ ($125$~mm pitch) and $83\%$ (design~A), the SHELL181 procedure gap widens monotonically from $2.4\%$ to $9.7\%$ and $14.0\%$.

Finally, restricting Eq.~\eqref{eq:rd} to junction elements yields the junction drilling ratio $r_{\mathrm{d},J}$. This ratio is insensitive to whether drilling is regularized using grounded springs or Hughes--Brezzi variational penalties, remaining $0.7977$ on design~A and $0.6439$ on the $250$~mm-pitch conforming panel. By contrast, the corresponding global ratio $r_{\mathrm{d}}$ drops from $1.104$ to $0.734$ and from $0.914$ to $0.675$, respectively. No spurious drilling modes occur below twice the critical load factor under in-plane shear or axial loading.
Known limitations of the baseline element under out-of-plane transverse edge loading are discussed in Section~\ref{sec:limitations}.

\section{Discussion}
\label{sec:discussion}

\subsection{Implications for engineering verification and design}

The primary practical implication of this study concerns the verification of finite element buckling analyses.
On cantilevered stiffened panels under in-plane bending and shear, commercial shell formulations produce diverging buckling predictions on both sides of the 3D continuum benchmark. Matrix export shows that ANSYS SHELL181 overpredicts through its uncoupled or membrane operator, whereas ANSYS SHELL281 shares the sub-block structure of the analytical full form. Indirect evidence indicates that Abaqus S4 behaves consistently with that group, while the substantially lower loads from Abaqus S8R are associated with differences in its internal stress stiffness assembly.
Consequently, matching an in-house solver or design code against a single commercial element establishes only which stress stiffness truncation the two share, not physical buckling capacity.

Mode-shape agreement alone is equally inconclusive on moderately stiffened panels, where distinct operators share nearly identical skin modes ($\mathrm{MAC} > 0.999$ on the $250$~mm-pitch panel). On densely stiffened panels, by contrast, omitting shear coupling changes the critical mode from a localized corner lift to a torsional twist ($\mathrm{MAC} = 0.574$ against S4; Fig.~\ref{fig:modes}).

Table~\ref{tab:activation} summarizes the three governing criteria of Section~\ref{sec:res-when} as a verification matrix across the benchmarked thin-walled configurations. The analytical-operator spread measures the maximum relative spread across the three analytical forms, $\max_{i,j}(\lambda_i/\lambda_j - 1)$ (reducing to $|\lambda_{\mathrm{block}}/\lambda_{\mathrm{full}} - 1|$ on the shear panels). The within-element procedure gap measures $|\lambda_{\mathrm{classic}} - \lambda_{\mathrm{pert}}|/\lambda_{\mathrm{classic}}$, while cross-element differences across distinct element families are considered separately.
Whenever any single criterion is absent---non-coplanar junctions, non-negligible junction bending or transverse shear, or junction modal drilling---the rotational stress stiffness terms contribute negligibly. In these cases, the three analytical operators coincide and within-element procedures remain closely aligned across the four six-DOF commercial elements (Table~\ref{tab:activation}). Remaining cross-element differences (typically $<0.8\%$, reaching $2.8\%$--$7.1\%$ on coarser channel, cylindrical, and shear-flow meshes) reflect interpolation, quadrature, and other element-level differences.

When all three criteria hold simultaneously on cantilevered shear panels, both the analytical-operator spread and the SHELL181 procedure gap grow markedly with junction density (Table~\ref{tab:activation}). On the converged level-3 mesh of shear design~A, the analytical spread reaches $34.4\%$ and the SHELL181 procedure gap reaches $13.4\%$. Across commercial element families on this mesh, predictions span $20.1\%$ below (Abaqus S8R) to $28.5\%$ above (classic SHELL181) the fine continuum reference.
Although optimal rib layouts remain robust across operators (Section~\ref{sec:res-continuum}), absolute buckling constraints and stability margins depend strongly on operator choice. Practitioners evaluating or optimizing densely stiffened shear panels should therefore cross-verify buckling loads using an alternative formulation (such as Abaqus S4 or ANSYS SHELL281) or a converged 3D continuum model rather than a single shell procedure.

\begin{table}[pos=tbp]
\centering
\footnotesize
\setlength{\tabcolsep}{3.0pt}
\caption{Mechanistic activation matrix for stress stiffness discrepancies across the benchmarked thin-walled configurations (spread and gap defined in Section~\ref{sec:discussion}).}
\label{tab:activation}
\begin{tabular}{>{\raggedright\arraybackslash}p{3.4cm}ccccc>{\raggedright\arraybackslash}p{4.0cm}}
\toprule
\makecell[l]{\textbf{Configuration}\\\textbf{(loading)}} &
\textbf{Junction} &
\makecell{\textbf{Jct}\\$M, Q$} &
\makecell{\textbf{Mode}\\$\theta_z$} &
\makecell{\textbf{Analyt.}\\\textbf{spread}} &
\makecell{\textbf{Proc.}\\\textbf{gap}} &
\makecell[l]{\textbf{Verification}\\\textbf{guidance}} \\
\midrule
Plain cylinder (axial, torsion) & No & -- & -- & $<0.1\%$ & $<0.1\%$ & No procedure sensitivity; standard mesh/element checks still apply. \\[2pt]
Stiffened panel (axial), stiffened cylinder, I-beam, box girder & Yes & Negligible & Varies & $<0.1\%$ & $<0.1\%$ & Membrane-dominated; single procedure adequate for operator screening. \\[2pt]
Stiffened panel (pure shear flow) & Yes & Yes & Weak & $0.12\%$ & $<0.1\%$ & Local skin lobing; junctions inactive (standard element checks apply). \\[2pt]
Channel cantilever (tip shear) & Yes & Mod. & Weak & $0.03\%$ & $0.27\%$ & Minor procedure sensitivity ($<0.3\%$); standard element checks apply. \\[2pt]
Conventional panels ($250$/$125$~mm, shear) & Yes & Yes & Yes & $4.8\%$ / $22.2\%$ & $2.4\%$ / $9.7\%$ & Cross-verify dense grids ($125$~mm pitch). \\[2pt]
\textbf{Dense panel~A (in-plane bending \& shear)} & \textbf{Yes} & \textbf{Yes} & \textbf{Yes} & \textbf{34.4\%--38.9\%} & \textbf{13.4\%--14.0\%} & \textbf{Cross-verify with S4, SHELL281, or 3D solid.} \\
\bottomrule
\end{tabular}
\end{table}
\FloatBarrier

\subsection{Methodological limitations and scope}
\label{sec:limitations}

The findings and methodology of this study are subject to four specific limitations:
\begin{enumerate}
  \item \textbf{Commercial classification evidence:} The stress stiffness operators of ANSYS SHELL181 and SHELL281 are established directly via exported system matrices and selective block elimination. In contrast, the classification of Abaqus S4 and S8R is inferred indirectly from load factors, critical mode shapes ($\mathrm{MAC} = 1.0000$ and $0.991$), and modal strain energies, because Abaqus does not export internal shell operators.
  \item \textbf{Continuum reference scope:} The baseline 3D 20-node solid continuum reference aligns with the \{Full form, S4, SHELL281\} group rather than SHELL181 or S8R, yet cannot discriminate within that $2.3\%$-wide group. Calibrating junction geometry under axial compression leaves shear junction-idealization sensitivity unresolved ($\approx 2.5\%/\mathrm{mm}$ under shear; $2.7\%/\mathrm{mm}$ under axial compression), so the raised-rib model (Table~\ref{tab:ribfoot}) is retained as an explicit geometric bracket. Combined with the intrinsic finite-rotation parameterization non-uniqueness of linearized shell buckling ($\alpha_{\mathrm{rot}} \in \{0, 1/2, 1\}$; Sections~\ref{sec:buckling} and~\ref{sec:res-continuum}), this precludes claiming any member of the group as uniquely correct.
  \item \textbf{Linearized bifurcation versus collapse capacity:} Linearized eigenvalue buckling provides an elastic bifurcation indicator for preliminary design and topology optimization, whereas ultimate load-carrying capacity of thin-walled panels depends on post-buckling stiffness, geometric imperfections, and material yielding. Although Section~\ref{sec:res-continuum} verifies that operator separation persists along the geometrically nonlinear pre-buckling path, this study assesses linearized elastic operator consistency rather than elastoplastic collapse loads.
  \item \textbf{Implementation limitations of the in-house element:} Three residual issues remain specific to the baseline in-house element and do not alter the commercial-element classifications or 3D continuum benchmarking under in-plane shear and axial loading:
  \begin{enumerate}[label=(\alph*), topsep=2pt, itemsep=2pt]
    \item \emph{Spurious drilling modes under transverse edge shear:} Under out-of-plane transverse edge shear ($-F_z$) on the axial-design layout, the weak bending-scaled grounded drilling spring ($10^{-4}\,\tr[(t^{3}/12)\mat{D}]$ per element) cannot stabilize an indefinite rotational block under primary skin bending. This produces spurious pure-drilling modes ($r_{\mathrm{d}} \gg 100$) at $43\%$ (MITC4) to $45\%$ (reduced shear) below the membrane buckling load; no such modes appear in ANSYS SHELL181 or under in-plane shear and axial compression below $2\lambda_{\mathrm{cr}}$.
    \item \emph{Transverse-bending stiffness offset:} On the fully stiffened panel under the same transverse edge shear, the membrane and block forms coincide and the full-form second-variation identity holds to $10^{-13}$. Nevertheless, the baseline (MITC4) and Variant~1 (reduced-shear) elements lie $13\%$ and $19\%$ below SHELL181 due to an unlocated linear-stiffness discrepancy under transverse bending.
    \item \emph{Compatible-membrane shear residual:} On the tied shear panels, the compatible-membrane baseline sits $1.28\%$ to $1.68\%$ above SHELL181 versus a $1.97\%$ axial offset, leaving a $0.3$ to $0.7$ percentage-point residual. Adopting Variant~2 (incompatible membrane modes and Hughes--Brezzi penalty) reduces the shear design~A discrepancy to $0.57\%$ (classic vs.\ block) and $0.12\%$ (perturbation vs.\ membrane).
  \end{enumerate}
\end{enumerate}

\section{Conclusions}
\label{sec:conclusions}

This study quantified how stress stiffness operator choices govern linearized buckling predictions of built-up stiffened shell panels across analytical formulations, commercial shell elements, and 3D continuum benchmarks. The main conclusions are summarized below:
\begin{enumerate}
  \item Matrix extraction shows that ANSYS SHELL181 switches operators across procedures, assembling the rotational block $\mat{K}_{\sigma,\theta\theta}$ in classic linear buckling but omitting it ($\mat{K}_{\sigma,\theta\theta} \approx \mat{0}$) in perturbation. In contrast, ANSYS SHELL281 retains both the rotational and translation--rotation sub-blocks characteristic of the analytical full form ($\mat{K}_{\sigma,\theta\theta}$ and $\mat{K}_{\sigma,u\theta}$). Indirect load-factor, modal, and strain-energy evidence indicates that Abaqus S4 behaves consistently with this full-form group, whereas the substantially lower loads of S8R are associated with differences in its internal stress stiffness assembly.
  \item The tested cases indicate that substantial formulation divergence requires three conditions simultaneously: non-coplanar junctions, non-negligible junction bending or transverse shear, and modal drilling rotations at those junctions. When any condition is absent---as in unbranched shells, pure axial compression, or pure membrane shear flow---analytical operators and within-element procedures remain consistent. On cantilevered shear panels where all three hold, both the analytical-operator spread and the SHELL181 procedure gap increase markedly with junction density.
  \item Independent 20-node solid continuum models (ANSYS SOLID186 and Abaqus C3D20) agree closely and indicate that the remaining through-thickness discretization effect is minor relative to the shell-operator differences. For the baseline junction geometry, the continuum reference aligns with the \{full form, Abaqus S4, ANSYS SHELL281\} group rather than the stiffened classic SHELL181 or the markedly softer Abaqus S8R predictions. Nevertheless, sensitivity to rib--skin junction idealization prevents the continuum benchmark from discriminating among the closely spaced members within that group.
  \item Linear buckling assessments and layout optimization of densely stiffened panels under in-plane bending and shear should not rely on a single commercial shell procedure. Predictions should be cross-verified using Abaqus S4, ANSYS SHELL281, or a converged 3D continuum model.
\end{enumerate}

\section*{Data availability}

A reproducibility archive for the stiffened panels, cylinders, and open beams is deposited under the MIT licence on Zenodo (\texttt{doi:10.5281/zenodo.22734110}).
It provides the in-house shell code assembling all three operators, discrete segment tables, every ANSYS and Abaqus input deck and result file, and an index mapping every reported number to its source script and data.
The container-crane box girder is excluded from the archive, but its mesh and self-weight model reconstruct directly from the full specification in Section~\ref{sec:panels} and Appendix~\ref{app:repro}.

\appendix

\section{Discrete rotational block at element level}
\label{app:discrete}

Because the transverse shear terms in Eq.~\eqref{eq:block} cannot be evaluated pointwise, Appendix~\ref{app:discrete} provides the discrete element matrix of the rotational block for the four-node shell element of Section~\ref{sec:femodel} in local coordinates.
Evaluating transverse shear terms pointwise introduces errors of tens of percent, whereas taking the frozen shear force at the element center misses the assembled rotational block by $13\%$ on pure rotational modes (Section~\ref{sec:verif:identity}).

The moment-weighted contribution integrates over $2\times2$ Gauss quadrature points $q$ with parent weights $w_q$ and Jacobian determinants $j_q$ (so $w_q j_q$ equals the combined weight denoted $w_q$ in Section~\ref{sec:buckling}).
Let $N_{a}$ denote the bilinear shape functions, $\mat{B}_{\mathrm{b}}$ the curvature--displacement operator generating pre-stress moments $\mat{M} = (t^{3}/12)\mat{D}\mat{B}_{\mathrm{b}}\vect{u}$, and $\mat{B}_{\mathrm{s}}(q)$ the MITC4 transverse shear operator generating $\mat{Q}(q) = (Q_x(q), Q_y(q))\T = k_{\mathrm{s}} G t\,\mat{B}_{\mathrm{s}}(q)\vect{u}$.
Expanding the derivatives $(\theta_z\theta_\alpha)_{,\beta}$ in Eq.~\eqref{eq:block} pairs the drilling rotation of node $a$ with the bending rotations of node $b$ via the symmetric gradient forms $S^{\alpha}_{ab} = N_{a,\alpha} N_b + N_a N_{b,\alpha}$:
\begin{equation}
  \mat{B}^{M}_{ab} = \tfrac{1}{2}\sum_q w_q j_q\,
  \Big[\, \big(M_{xx} S^{x}_{ab} + M_{xy} S^{y}_{ab}\big)(q)\,\mat{e}_x
        + \big(M_{yy} S^{y}_{ab} + M_{xy} S^{x}_{ab}\big)(q)\,\mat{e}_y \,\Big],
  \label{eq:blockM}
\end{equation}
where $\mat{e}_\alpha$ places each scalar into the $(\theta_z, \theta_\alpha)$ entry of the $3\times3$ rotational sub-block for node pair $(a,b)$.
The resulting matrix is symmetrized (canceling the leading $1/2$ factor) so that $\vect{\varphi}\T\Kg\vect{\varphi}$ evaluates Eq.~\eqref{eq:block} exactly on mode $\vect{\varphi}$.
Pre-stress moments $M_{\alpha\beta}$ are evaluated at the same Gauss points using $\mat{B}_{\mathrm{b}}$.

In contrast, the transverse shear contribution must be assembled at the four MITC4 edge-midpoint tying points $p = 1, \ldots, 4$ used by the linear stiffness matrix.
Here $H_{qp}$ interpolates covariant tying strains to Gauss point $q$. The map $d(p) \in \{1,2\}$ assigns each tying point $p$ to its associated covariant coordinate direction. The $2\times 2$ inverse Jacobian $\mat{T}_q = [T_{i,d}(q)]_{i,d\in\{1,2\}}$ maps covariant direction $d$ to Cartesian component $i \in \{1,2\}$; thus, entry $T_{i,d(p)}(q)$ maps tying point $p$ into Cartesian component $i$.
Matrices $\mat{G}^{\mathrm{t}}_{p}$ and $\mat{B}^{\mathrm{t}}_{p}$ form the covariant tying strain $\gamma_p$ from nodal deflections $\vect{w}$ and rotations $\vect{\theta}$. Matrix $\mat{N}^{\mathrm{t}}_{p}$ holds the nodal shape functions $N_a$ evaluated at tying point $p$. The quantities $(b^{x}_{p}, b^{y}_{p})$ are the Cartesian components of the local covariant tangent vector along the edge associated with tying point $p$; consequently, $\mat{B}^{\mathrm{t}}_{p}\vect{\theta} = b^{x}_{p}\theta_y - b^{y}_{p}\theta_x$ at point $p$.

The frozen shear force at Gauss point $q$ follows directly from the assumed tying field:
\begin{equation}
  \begin{pmatrix} Q_x(q) \\ Q_y(q) \end{pmatrix}
  = k_{\mathrm{s}} G t \,
    \mat{T}_q \begin{pmatrix} \sum_{p:\,d(p)=1} H_{qp}\, \gamma_p \\[2pt]
                              \sum_{p:\,d(p)=2} H_{qp}\, \gamma_p \end{pmatrix}
  = k_{\mathrm{s}} G t\,\mat{B}_{\mathrm{s}}(q)\vect{u},
  \qquad
  \gamma_p = \mat{G}^{\mathrm{t}}_{p}\vect{w} + \mat{B}^{\mathrm{t}}_{p}\vect{\theta},
  \label{eq:frozenQ}
\end{equation}
where each sum runs over the two tying points associated with its covariant coordinate direction.
Integrating that force back against the interpolation functions yields the scalar weight $c_p$ at each tying point:
\begin{equation}
  c_p = \sum_q w_q j_q\, H_{qp}\,
        \big[\, T_{1,d(p)}(q)\, Q_x(q) + T_{2,d(p)}(q)\, Q_y(q) \,\big].
  \label{eq:tieweight}
\end{equation}
The shear-weighted contribution to the rotational block then becomes $\mat{B}^{Q}_{ab} = \tfrac{1}{2}\sum_{p=1}^{4} c_p\, \big[\mat{N}^{\mathrm{t}}_{p}\big]_a \big[\mat{N}^{\mathrm{t}}_{p}\big]_b \big(\,b^{x}_{p}\,\mat{e}_x + b^{y}_{p}\,\mat{e}_y\,\big)$, inserted into the $(\theta_z, \theta_\alpha)$ entries and symmetrized.
Any implementation forming Eq.~\eqref{eq:frozenQ} at the element center or replacing Eq.~\eqref{eq:tieweight} by pointwise Gauss-point sampling fails the energy identity in Section~\ref{sec:verif:identity}, making Eq.~\eqref{eq:identity} a sensitive diagnostic of discrete shear consistency.

\section{Adjoint load of the rotational block}
\label{app:adjoint}

Section~\ref{sec:sensitivity} summarizes the eigenvalue sensitivity when pre-stress depends on design variables.
For the membrane form, the adjoint load is a stress-like vector quadratic in the buckling mode $\vect{\varphi}$ and applied through the membrane strain operator.
Naive differentiation that swaps the equilibrium state and mode in Eq.~\eqref{eq:kge} assumes a false symmetry in the trilinear form and fails finite-difference verification by orders of magnitude.
Below, we derive the corresponding adjoint loads for the rotational block of Eq.~\eqref{eq:block} using the element operators defined in Appendix~\ref{app:discrete}.

Because each sub-block of $\Kg(\vect{u})$ is linear in the displacement state $\vect{u}$ via a single resultant, differentiating $\vect{\varphi}\T\Kg(\vect{u})\vect{\varphi}$ with respect to $\vect{u}$ pairs a quadratic modal vector with the corresponding linear strain-displacement operator.
In the local element frame, let $\theta_x$ and $\theta_y$ denote nodal bending rotations of $\vect{\varphi}$, $\theta_z$ denote its drilling rotation, and $\mat{B}_{\mathrm{b}}$ and $\mat{B}_{\mathrm{s}}(q)$ denote the curvature and MITC4 transverse shear operators of Appendix~\ref{app:discrete}.

Using $\mat{M} = (t^{3}/12)\,\mat{D}\,\mat{B}_{\mathrm{b}}\vect{u}$, the moment-weighted rotational block pairs modal rotation-product gradients $\vect{\pi}_q$ with $\mat{B}_{\mathrm{b}}$ at each Gauss point $q$:
\begin{equation}
  \sum_q w_q j_q\,\mat{B}_{\mathrm{b}}\T\,\frac{t^{3}}{12}\,\mat{D}\,\vect{\pi}_q ,
  \qquad
  \vect{\pi}_q = \begin{bmatrix}
    (\theta_z\theta_x)_{,x} \\[2pt]
    (\theta_z\theta_y)_{,y} \\[2pt]
    (\theta_z\theta_x)_{,y} + (\theta_z\theta_y)_{,x}
  \end{bmatrix} ,
  \label{eq:adjrot}
\end{equation}
matching the three resultant contractions in Eq.~\eqref{eq:block}.

For the transverse shear contribution, where $\mat{Q}(q) = k_{\mathrm{s}} G t\,\mat{B}_{\mathrm{s}}(q)\vect{u}$, pointwise Gauss sampling is invalid.
Instead, differentiating Eq.~\eqref{eq:tieweight} pairs the quadratic modal rotation products $r_p$ at the four MITC4 edge-midpoint tying points $p$ with the assumed-strain shear operator at each Gauss point $q$:
\begin{equation}
  \sum_{q} w_q j_q\,\mat{B}_{\mathrm{s}}(q)\T\,k_{\mathrm{s}} G t \sum_{p=1}^{4} H_{qp}\, r_p
  \begin{pmatrix} T_{1,d(p)}(q) \\[2pt] T_{2,d(p)}(q) \end{pmatrix} ,
  \qquad
  r_p = \left.\big(b^{x}_{p}\,\theta_z\theta_x + b^{y}_{p}\,\theta_z\theta_y\big)\right|_{p} ,
  \label{eq:adjshear}
\end{equation}
using $H_{qp}$, $T_{i,d(p)}(q)$, and $(b^{x}_{p}, b^{y}_{p})$ from Appendix~\ref{app:discrete}.
For Variant~1 (one-point reduced shear), replacing $\mat{B}_{\mathrm{s}}(q)$ by the constant element-center shear operator recovers the reduced-integration adjoint directly, whereas evaluating $r_p$ directly at Gauss points violates the energy identity of Section~\ref{sec:verif:identity} just as observed for the forward rotational block.

Equations~\eqref{eq:adjrot} and~\eqref{eq:adjshear} provide a direct internal verification check independent of finite differences: neither adjoint load acts on the drilling DOF.
Although mode products include $\theta_z$, the equilibrium curvature $\kappa$ and shear strain $\gamma$ depend only on translations and bending rotations; any non-zero drilling component in the adjoint load therefore reveals differentiation with respect to the wrong argument.
Adjoint loads for the translation--rotation couplings of Eq.~\eqref{eq:couplings} in the full form follow analogously by differentiating Eq.~\eqref{eq:couplings} with respect to $\vect{u}$ and are included in the reproducibility archive. Both the block form and the full form pass an end-to-end central finite-difference check (step $10^{-6}$ across six element densities) over static solve, matrix assembly, and eigensolution on a $12 \times 4$ panel.

\section{Reproducibility and solver specifications}
\label{app:repro}

\subsection{In-house element stabilization and variants}

Parameter sweeps fixed two stabilization choices in the in-house element of Section~\ref{sec:femodel} and Table~\ref{tab:variants}.
First, the reference stiffness $\Kref$ used by multipoint constraints and rotational regularization is taken strictly over skin DOFs rather than globally.
Taking a global maximum tracks the most distorted rib element: on one graded panel, this inflated the drilling spring to $15\%$ of skin bending stiffness and raised the predicted load factor to $8.71$ versus SHELL181's $4.53$.

Second, sweeping the rotational stabilization parameter of Eq.~\eqref{eq:rotreg} over $\varepsilon_r \in [10^{-9}, 10^{-13}]$ on an optimized panel relative to SHELL181 yielded load factors $10.0\%$, $2.4\%$, $1.4\%$, $1.3\%$, and $1.3\%$ above SHELL181 at $\varepsilon_r = 10^{-9}, 10^{-10}, 10^{-11}, 10^{-12}$, and $10^{-13}$.
The value $\varepsilon_r = 10^{-11}$ is retained throughout, lying two orders of magnitude above the conditioning floor while differing from the converged limit by $\sim 0.1\%$.

The baseline four-node element employs the MITC4 shear field~\citep{Bathe1985a} consistently in the stiffness matrix, second-order stress stiffness strain, and frozen shear resultants under $2\times2$ Gauss integration, paired with the bending-scaled drilling spring and $\varepsilon_r = 10^{-11}$ (Section~\ref{sec:femodel}).
Variant~1 retains this formulation while switching both linear transverse shear stiffness and frozen shear resultants to one-point central Gauss integration.
Variant~2 enriches the membrane field with four condensed QM6 incompatible modes~\citep{Taylor1976} using centroid-Jacobian derivative scaling so distorted quads pass the constant-strain patch test, retains MITC4 transverse shear, and replaces the grounded drilling spring with the Hughes--Brezzi variational drilling penalty~\citep{Hughes1989a} at dimensionless factor $\gamma_{\mathrm{d}} = 1$ evaluated at the element center.
For all optimized panels, discrete active element sets are selected using the minimum density of each segment's two endpoints so that exported commercial decks match the in-house geometry identically.

\subsection{Commercial shell and 3D continuum solver settings}

ANSYS analyses were run in batch (four processes) on Mechanical APDL Release~19.2 and verified on Release~2024~R2, reproducing every load factor across both procedures, shell elements, and exported matrices within $0.0003\%$.
Shell models use SHELL181 (\deck{KEYOPT(3)}~$= 2$, \deck{KEYOPT(8)}~$= 0$) or SHELL281 with default drilling settings; tied models apply constraint equations across all six DOFs, whereas shared-node rebuilds and new structures omit ties.

In ANSYS, the classic linear eigenvalue procedure executes a linear static solve (\deck{PSTRES,ON}) followed by \deck{ANTYPE,BUCKLE} and \deck{BUCOPT,LANB}.
The perturbation procedure performs a nonlinear static run (\deck{NLGEOM,ON}, five substeps, \deck{RESCONTROL,LINEAR,ALL,1}) at base load $\alpha = 0.01$ (or $0.1$ to $8.6$ along the base-load sweep of Fig.~\ref{fig:ladder}), followed by \deck{ANTYPE,STATIC,RESTART,,,PERTURB}, \deck{PERTURB,BUCKLE}, and \deck{SOLVE,ELFORM} under unit reference load.
The signed total load factor is $\alpha + \beta$ (with $\beta$ read from the \deck{LOAD MULTIPLIER} output block), and the reported critical load-factor magnitude is $\lambda_{\mathrm{cr}} = |\alpha + \beta|$ (Section~\ref{sec:coverage}).
Sequential geometrically nonlinear static solves (\deck{NLGEOM}) followed by perturbation solves are also used along the reversed-load equilibrium paths in Section~\ref{sec:res-continuum}.

Global stiffness (\deck{STIFF}) and stress stiffness (\deck{MASS}) matrices are extracted via \deck{*SMAT} (\deck{IMPORT,FULL}) and written in Matrix Market format using \deck{*EXPORT}, with row indices identified from \deck{HBMAT} mapping files in \deck{/AUX2}; when absent, rotational rows sit cleanly below the $60$-to-$650$-fold stiffness-diagonal gap and match the \deck{HBMAT} mapping identically.
ANSYS stores the stress stiffness matrix with the opposite sign convention to Eq.~\eqref{eq:eigproblem}.

Abaqus/Standard~2026 analyses use S4, S4R, S8R, or S8R5 with five section integration points.
Rib--skin ties use \deck{*EQUATION} constraints eliminating the rib node and omitting constrained boundary DOFs.
The classic procedure runs one Lanczos \deck{*BUCKLE} step with a negative lower eigenvalue bound to extract both spectrum branches.
The perturbation procedure runs a geometrically nonlinear \deck{*STATIC} step (\deck{NLGEOM}) at $0.01$ reference load followed by a \deck{*BUCKLE} step at unit load, yielding signed total load factor $0.01 + \beta$ and reported critical magnitude $\lambda_{\mathrm{cr}} = |0.01 + \beta|$ (Section~\ref{sec:coverage}); modes are written via \deck{*NODE PRINT} for \deck{U} and \deck{UR}.
Standard commercial keyword definitions follow the vendor theory documentation~\citep{AnsysTheory2024,AbaqusTheory2024}.

Continuum benchmarks use SOLID186 (\deck{KEYOPT(2)}~$= 1$) and Abaqus C3D20 on tensor-grid meshes aligned with every rib face, clamped at $x = 0$ and loaded by consistent end tractions on the skin face.
Across the coarse, medium, and fine baseline meshes, these grids place $2$, $3$, and $4$ elements between rib faces, $1$, $2$, and $2$ elements through wall thickness, and $4$, $6$, and $8$ elements over rib height.
The through-thickness refinement study of Table~\ref{tab:thickness} refines both skin and rib walls to $1$, $2$, and $4$ quadratic layers on the coarse in-plane grid and to $2$ and $4$ layers on the medium grid. Four-layer refinement was also attempted on the fine grid (\texttt{fine\_thk4}) but was terminated because its memory and scratch requirements exceeded the 64~GB workstation limits.
For the rib-height sensitivity bracket in Table~\ref{tab:ribfoot}, raising the rib top by $5$~mm exposes a full $100$~mm standing blade above the skin top face.

\subsection{Panel topology counts and box-girder reconstruction}

Detailed topology counts for the stiffened panels and complete reconstruction parameters for the container-crane box girder complement Table~\ref{tab:structures}.
On the tied $48\times 16$ lattice ($n_z = 3$ rib-height divisions, $n_{\mathrm{sub}} = 2$ skin subdivisions per cell, $10\,368$ candidate elements, and $9\,732$ nodes), shear design~A has $641$ active segments ($40.06$~m total length) joining $383$ nodes ($161$ crossings, $198$ T-junctions, $17$ corners, $4$ free ends, and $3$ straight continuations).
Shear design~B, the combined design, and the axial design each have $640$ segments, with crossing/free-end/T-junction counts of $164$/$2$/$\ge 190$ (design~B), $164$/$5$/$\ge 190$ (combined), and $109$/$8$/$90$ (axial).
The $250$~mm-pitch conforming panel has $448$ segments ($28.00$~m) with $33$ crossings, $28$ boundary T-junctions, and $0$ free ends, whereas the cut-rib variant introduces $36$ free ends.
In shared-node rebuilds (Table~\ref{tab:levels}), subdivision levels~1, 2, and~3 place $1$, $4$, and $9$ eight-node elements (or $4$, $16$, and $36$ four-node elements) per design cell on identical node lattices.

For the container-crane box girder (Fig.~\ref{fig:boxgirder}), the $14$~mm flanges overhang the $8$/$10$~mm webs by $0.15$~m, while $34$ interior $8$~mm diaphragms at $1.0$~m pitch feature $0.8\times 1.8$~m central manholes alongside solid end plates.
The $333\times 24\times 44$ shell grid ($67\,716$ elements, MITC4 transverse shear in both $\mat{K}$ and $\Kg$) places $315$ cells between bearing centerlines and $9$ cells on each $1.0$~m end overhang.
At the two $2.0$~m bearing pads, lateral displacement is restrained across each pad, vertical displacement and transverse tilt are held along the pad centerline, and longitudinal translation is restrained at the left pad only.
Loading combines two $200$~kN wheel loads spaced $9.0$~m apart over the $10$~mm web with $261$~kN of self-weight ($\rho = 7850$~kg/m$^{3}$, $g = 9.81$~m/s$^{2}$).

Table~\ref{tab:when} compiles the supplementary pre-stress ratios, modal drilling ratios, and membrane- versus block-form load factors for the tied optimization panels and the container-crane box girder, confirming that the box girder remains unactivated ($r_{\mathrm{d}} = 0.0007$, zero block shift at $\lambda = 1.7223$).

\begin{table}[pos=!htbp]
\centering
\caption{Pre-stress resultant ratios, mode drilling ratios, and membrane- versus block-form load factors on tied optimization panels and the box girder.}
\label{tab:when}
\begin{tabular}{lrrrrr}
\toprule
 & \multicolumn{2}{c}{Pre-stress ratios} & Mode & \multicolumn{2}{c}{Buckling load factor} \\
\cmidrule(lr){2-3}\cmidrule(lr){4-4}\cmidrule(lr){5-6}
Configuration & $6\bar M/(\bar N t)$ & $Q/\bar N$ & $r_{\mathrm{d}}$ & Without block & With block \\
\midrule
axial design    & $0.0154$ & $0.00073$ & $1.6469$ & $10.9121$ & $10.9121$ \\
shear design A  & $1.1750$ & $0.10488$ & $1.3677$ & $8.7003$  & $10.2577$ \\
combined design & $1.1201$ & $0.09923$ & $1.3654$ & $5.8700$  & $6.7411$  \\
conforming panel$^{\dagger}$ & $1.1599$ & $0.04491$ & $0.7834$ & $4.7765$  & $4.9081$ \\
box girder      & $0.0144$ & $0.00027$ & $0.0007$ & $1.7223$  & $1.7223$  \\
\bottomrule
\end{tabular}
\\[0.5ex]{\footnotesize $^{\dagger}$ Evaluated under Variant~2 of Section~\ref{sec:femodel}.}
\end{table}

\end{document}